\documentclass[preprint,article,accept,moreauthors,pdftex]{Definitions/mdpi_pre} 

\firstpage{1} 
\pubvolume{xx}
\issuenum{1}
\articlenumber{preprint} 
\nameofjournal{journal} 

\pubyear{2022}
\copyrightyear{2022}
\externaleditor{} 
\history{} 

\setcitestyle{authoryear}
\usepackage{subcaption}
\usepackage{wrapfig}
\renewcommand*\textcircled[1]{\tikz[baseline=(char.base)]{
            \node[shape=circle,draw=black,inner sep=0pt, minimum size = .4cm] (char) {#1};}
            }
\newcommand*\tcw[1]{\tikz[baseline=(char.base)]{
            \node[shape=circle,draw=white,inner sep=0pt, minimum size = .4cm] (char) {#1};}
            }

\Title{A combined optimization–simulation approach for modified outside-in boarding under COVID-19 regulations including limited baggage compartment capacities}

\Author{Michael Schultz $^{1}$, Majid Soolaki $^{2}$, Mostafa Salari$^{3}$, and Elnaz Bakhshian$^{4}$}
\AuthorNames{Michael Schultz, Majid Soolaki, Mostafa Salari, Elnaz Bakhshian}

\address{%
$^{1}$ \quad Institute of Flight Systems, Bundeswehr University Munich, 85577 Neubiberg, Germany\\
$^{2}$ \quad School of Architecture and Cities, University of Westminster, London, United Kingdom\\
$^{3}$ \quad College of Engineering and Mines, Department of Civil Engineering, University of North Dakota\\
$^{4}$ \quad School of Civil Engineering, University College Dublin}

\firstnote{The authors contributed equally to this work.} 

\abstract{The timely handling of passengers is critical to efficient airport and airline operations. The pandemic requirements mandate adapted process designs and handling procedures to maintain and improve operational performance. Passenger activities in the confined aircraft cabin must be evaluated to potential virus transmission, and boarding procedures should be designed to minimize the negative impact on passengers and operations. In our approach, we generate an optimized seat allocation that considers passengers' physical activities when they store their hand luggage items in the overhead compartment. We proposed a mixed-integer programming formulation including the concept of shedding rates to determine and minimize the risk of virus transmission by solving the NP-hard seat assignment problem. 
We are improving the already efficient outside-in boarding, where passengers in the window seat board first and passengers in the aisle seat board last, taking into account COVID-19 regulations and the limited capacity of overhead compartments.
To demonstrate and evaluate the improvements achieved in aircraft boarding, a stochastic agent-based model is used in which three operational scenarios with seat occupancy of 50\%, 66\%, and 80\% are implemented.
With our optimization approach, the average boarding time and the transmission risk are significantly reduced already for the general case, i.e., when no specific boarding order is specified (random boarding).
If the already efficient outside-in boarding is used as a reference, the boarding time can be reduced by more than 30\% by applying our approach, while keeping the transmission risk at the lowest level.}


\begin{document}


\section{Introduction}
\label{sec:DP}
Worldwide economic sectors were severely hit by the coronavirus outbreak (COVID-19). In the aviation sector, at the beginning of February 2020, the outbreak started to negatively impact the number of worldwide flights~\citep{dube2021COVID,ICAO2021}. International Civil Aviation Organization (ICAO) reported a 25.81 percent to 71.78 percent decrease in the number of flights from March 2020 to December 2020 compared to the last year~\citep{ICAO2021}. This unprecedented decline in the number of flights is because flights require many passengers to spend time in the aircraft cabin, which exposes them to a high risk of infection. This considerable chance of infection along with the notorious history of the aviation industry in previous pandemics, i.e., MERS-CoV, reduced the willingness of passengers to involve in air travels in the current outbreak~\citep{forsyth2020COVID,iacus2020estimating,regan2016tracing}. Airlines are required to take prompt action to accommodate the outbreak situation to ensure passengers (health safety~\citep{serrano2020future}). The international air transportation association (IATA) medical advisory group suggests social distancing measures, also introduced as physical distancing, among passengers to reduce the chance of infection. This measure includes sufficient distancing among passengers while seated, during check-in, and through the boarding process \citep{iata2020restarting}. World health organization (WHO) and European Union Aviation Safety Agency (EASA) also emphasize the importance of social distancing measures to reduce the rate of infections among people who share the same activities \citep{who2020, EASA2020}. According to the instruction published by EASA, “airplane operators should ensure, to the extent possible, physical distancing among passengers”. Some recent studies stressed that the ideal practice of social distancing among passengers requires a well-defined seating assignment methodology as well as the willingness of passengers to practice social distancing guidelines \citep{pavlik2021airplane,Salari2020_101915}.

\begin{figure}[htb!]
    \centering
    \includegraphics[width=.6\textwidth]{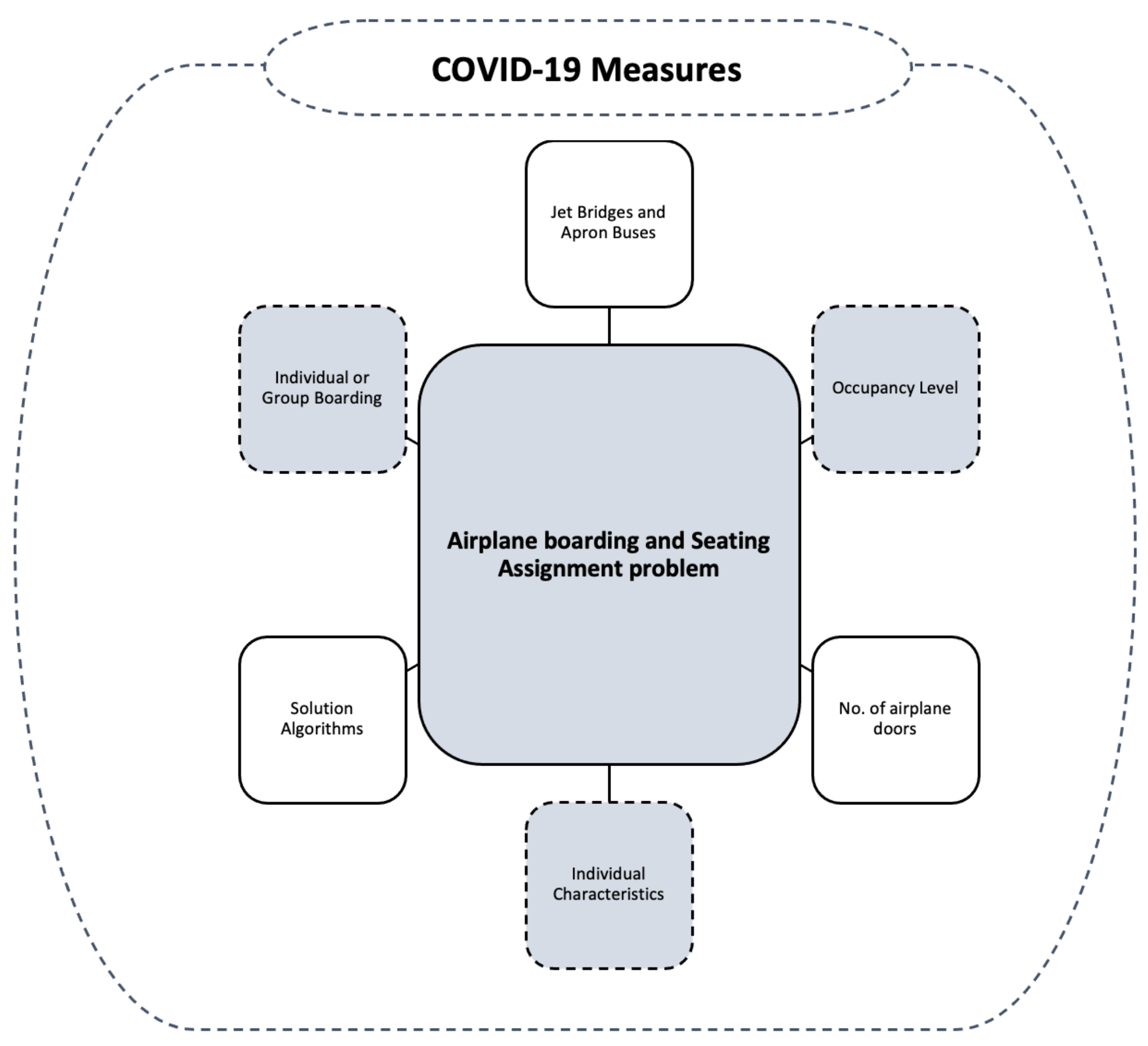}
    \caption{Different aspects of Airplane Boarding and Seating Assignment Problem with the consideration of the COVID-19 measures}
    \label{fig:FIG_1}
\end{figure}

Fig.~\ref{fig:FIG_1} illustrates the main considerations for the airplane boarding and seating assignment problem while the studies focusing on this problem can be similarly categorized. The involvement of COVID-19 measures in this problem primarily focuses on some of the considerations including the boarding strategies (individual, group), level of seat occupancy, and individual passenger characteristics. 
Concerning the individual characteristics, there are studies that incorporate the passengers’ carry-on bags into the aircraft boarding and seating assignment problem without the consideration of COVID-19 measures as the number of bags can affect the walking speed of passengers and luggage storage time and therefore affect the overall boarding time ~\citep{milne_new_2014,MILNE2016_104,schultz_implementation_2018,qiang2014reducing}. For example, if passengers' luggage is distributed evenly in the cabin, boarding time can also be reduced~\citep{milne_new_2014}.

With respect to the COVID-19 measures, physical activities with higher intensity, such as storing luggage, may increase the risk of virus transmission to passengers who are at a close distance \citep{schultz_evaluation_2020}. Therefore, while missing in the current literature, the consideration of passengers' carry-on luggage can assist in an optimized assignment of passengers to seats to reduce the chance of virus transmission among passengers. The gap in previous work to analyze the effect of carry-on bags on the physical distancing of passengers in seat allocation and boarding sequences is our primary research focus for this work.

Our research makes two new and important contributions. First, we provide a theoretical approach by developing a model that accounts for seat assignment and passenger boarding sequence, considering the utilization of the capacity-limited overhead compartment. This is a significant contribution, as recent studies emphasize the relevance of physical interactions for the virus transmission (cf.~\cite{schultz_evaluation_2020,salari2021social,Michael_Majid_TRC_2021, Michael_Majid_TransportB_2021}). We focus on outside-in boarding sequence, which has already been evaluated as very efficient since it eliminates unfavorable passenger interactions within the seat rows (window seat is boarded first, followed by middle seat and aisle seat). However, we will also show that our approach leads to a significant improvement in general cases (random boarding sequence).

As the overhead compartments become more utilized during boarding, passengers must put in the additional physical effort to first rearrange the luggage in the compartment to make room for their own luggage. Second, with respect to a methodological contribution, we researched the combination of optimization and simulation approaches to handle the aircraft boarding problem under COVID-19 requirements, providing the optimized seat allocations and boarding sequences. Here, the proposed mixed-integer programming (MIP) optimization problem provides a solution for the passenger seat assignment. Given the high complexity of the MIP problem, we develop a genetic algorithm to obtain near-optimal solutions. Finally, the stochastic and agent-based simulation approach determines and evaluates appropriate passenger boarding sequences based on the given seat assignment.

The paper is structured as follows. After providing a relevant background in the literature review (Section~\ref{sec:literature}), we introduce our optimization model for the seat assignment problem in Section~\ref{sec:optimization} assuming an outside-in boarding sequence, which already reduces additional passenger interaction during seating. In Section~\ref{sec:optresults}, we design a genetic algorithm to solve the NP-hard seat assignment problem. Therefore, we provide operators for mutation, crossover, migration, and elitism and solve three use cases with different seat loads. The resulting seat allocations are input for the following agent-based simulation approach, which is used to determine the appropriate boarding sequence (Section~\ref{sec:agentSimulation}), considering passenger interactions, virus transmissions risk, and the increased utilization of the overhead during the boarding progress. In our approach, we have not yet implemented a feedback loop from the sequence optimization to the seat assignment model. This will be part of future research work. Finally, our contribution ends with a conclusion and outlook in Section~\ref{sec:discussion} .

\section{Literature review}
\label{sec:literature}
The brief literature review is organized in three subsections, with a focus on common optimization fields, aircraft boarding, and particular challenges related to pandemic requirements (e.g., physical distance).

\subsection{Optimization strategies}

The development of mathematical modelling and optimization approaches has been suggested by many researchers in different areas such as production planning \citep{Pochet_2006, msly_2012, Bersch_2021}, location allocation problems \citep{Cooper_1963, Ishfaq_2011, Cheng_2021},  supply chain designing \citep{Lopez_2003, SoolakiArkat_2018, Polo_2019}, evacuation planning \citep{KONGSOMSAKSAKUL_2005, Shahparvari_2017, Bakhshian_2021}, vehicle routing problems \citep{Laporte_1992, Baker_2003, Alinaghian_2018}, air transportation problems \citep{Bertsimas_2011, Das_2020, Yilmaz_2021}, and boarding and disembarking problems \citep{majid_soolaki_2012,MILNE2016_104, Michael_Majid_TRC_2021,Michael_Majid_TransportB_2021,Michael_Majid_ATM_2021,Xie_deboarding_2021}. 

A variety of techniques are proposed to address aircraft boarding problem, such as linear and mixed-integer programming (MIP)~\citep{bazargan2007linear,milne2018robust,MILNE2016_104,salari2019airplane}, meta-heuristic algorithms \citep{majid_soolaki_2012,wittmann2019customer}, discrete simulation \citep{milne_new_2014,steffen2008statistical,tang2012aircraft,van_den_briel_america_2005}, grid-based simulation \citep{Schultz_2017_dynamic_change, zeineddine2021reducing}, agent-based and stochastic modeling \citep{milne2019greedy,schultz_fieldTrial_2018,schultz_implementation_2018}; and empirical experiments \citep{steffen_experimental_2012}.

\subsection{Aircraft boarding}
The aircraft boarding studies focuses on minimizing the boarding time of passengers to decrease airline operation cost incurred by aircraft turnaround time~\citep{schultz_future_2020,hutter2019influencing}. This research direction can be classified concerning the boarding assumptions and modeling techniques. There are studies that assume that the jet bridges are used to transfer passengers from the boarding gate to aircraft~\citep{bachmat2009analysis,jaehn2015airplane,milne2018robust,milne_new_2014,MILNE2016_104,salari2019airplane,schultz_fieldTrial_2018,schultz_implementation_2018} while others consider apron buses for passengers commuting to aircraft \citep{delcea2019methods,delcea2018two,milne2019new}. Studies address this problem:

\begin{enumerate}
    \item under different level of seat occupancy \citep{kierzkowski2017human,notomista2016fast,qiang2014reducing,steffen2008optimal,schultz_implementation_2018}
    \item with the assumption of passengers boarding through one door or both front and rears doors of an aircraft \citep{kuo2015improved,milne2019new,steiner2009speeding,schultz_efficiency_2008},
    \item concerning individual characteristics of passengers including walking time and number of carry-on bags \citep{hutter2019influencing,kierzkowski2017human,milne2018robust,schultz2018consideration},
    \item assuming passengers traveling in groups \citep{tang2019extended,tang2012aircraft,wittmann2019customer,Michael_Majid_TRC_2021}, and 
    \item considering seating assignment~\citep{2005_Ferrari_Robustness,salari2019airplane,steffen2008optimal}. 
\end{enumerate}

\subsection{Aviation management during coronavirus outbreak}
Researchers have been studying the impact of the corona\-virus pandemic from the beginning to better understand its consequences. Concerning coronavirus related studies in the aviation industry, the focus was set on the collapse in air travel demand and airport charges~\citep{forsyth2020COVID}, global airline industry~\citep{maneenop2020impacts} and airlines’ employment~\citep{Sobieralski_2020}, and estimation and projection of air traffic evolution and its socio-economic impact~\citep{iacus2020estimating}. The pandemic has significant implications for airport capacity and service levels~\citep{serrano2020future}, and in particular, for the future of aircraft handling operations due to (post-)pandemic requirements~ \citep{schultz_future_2020}. In this context, new technologies are needed to efficiently determine passenger locations in indoor environments and confined aircraft cabins~\citep{schwarzbach2020technology}. As physical distancing measure exhibits a great opportunity to reduce the spread of coronavirus among people \citep{sen2020social}, this measure has been studied in a broad range of scientific works including political, economic, and social challenges \citep{yezli2020COVID}, and ethical aspects of physical distancing \citep{lewnard2020scientific,Salari2020_101915}.

The pandemic requirements, in particular the requirement for sufficient distances between passengers or groups of passengers, have a lasting effect on the process flows and times for boarding and disembarking. The primary objective is to minimize the risk of transmission as far as possible and to develop appropriately adapted processes~\citep{schultz_evaluation_2020,milne2021airplane,Salari2020_101915}. To address the situation where passengers travel in groups, a new analytical approach was designed to optimize the seating layout of passengers to minimize the spread of virus \citep{Michael_Majid_TRC_2021}. The developed approach was also used by the authors to study an optimized passenger disembarkation process considering COVID-19 regulations~\citep{Michael_Majid_TransportB_2021}. In this context, the developed model not only optimizes the boarding and disembarkation time but also minimizes the risk of virus transmission.

\section{Optimization model}
\label{sec:optimization}

We build the optimization model for the passenger seating assignment problem based on an outside-in sequence of boarding in which the window-seat passengers board first. It is followed by the boarding of middle-seat passengers and aisle-seat passengers, respectively. We use outside-in here because the problem of additional interaction during seating, e.g., passengers must leave their seats when window seat passengers arrive last, is already solved. Note that the model can be easily adjusted to consider other strategies for boarding passengers, for instance, to an optimal individual, back-to-front staggered sequence of passengers, which is a border case of the outside-in boarding strategy~(cf.~\cite{steffen2008optimal,Schultz_2017_dynamic_change,Michael_Majid_TRC_2021}). 
 
Relaxing \emph{known boarding sequence} will add to the complexity of the formulation with little insight into the effect of passengers' carry-on bags on the virus spread. Therefore, we keep the assumption that the boarding sequence is known. This assumption is in line with \citep{MILNE2016_104} as they assume the sequence of boarding is known when they focus on seating assignment of passengers with carry-on bags. 

Based on the assumptions of the seating assignment problem, we list sets, parameters, and decision variables. We define the number of rows, columns, the number of carry-on bags, and interaction types as $I$, $J$, $B$, and $R$ respectively. Our use case is for a single-aisle aircraft with 174 seats, which is typical of most of the Airbus A320 and Boeing B737 family aircraft in service and which is used as a reference case in several research studies. The number of rows and columns are $I = 29$ and $J = 6$ and we assume that the number of bags for each passenger can range between zero and two. 

\noindent
\begin{table}[htb!]
    \centering
    \begin{tabular}{p{2.5cm} p{11cm}}
        \textbf{Notation} &  \textbf{Definition} \\
        \multicolumn{2}{l}{\emph{Sets and Indexes}}\\
        $i$ & Index set of row $i\in \{1,2, \dots, \mathcal{I}\}$\\
        $j$ & Index set of column $j\in \{1,2, \dots, \mathcal{J}\}$\\
        $b$ & Index set of the number of bags $b~\in~\{1,2, \dots,\mathcal{B}\}$\\
        $r$ & Index set of interaction type during the flight $r~\in~\{1,2, \dots, \mathcal{R}\}$\\
        $r'$ & Index set of interaction type during storing the bags $r'\in \{1,2, \dots, \mathcal{R'}\}$\\
        \\
        \multicolumn{2}{l}{\emph{Parameters}}\\
        $\textit{Bag}_{b}$ & Number of passengers who has $(b-1)$ bag(s)\\
        $\textit{SR}_{r}^\textit{flight}$ & Related flight shedding rate for interaction $r$ when passengers are seated during flight\\
        $\textit{SR}_{r'}^\textit{store}$ & Related storing shedding rate for interaction $r'$ when a passenger stores bag(s)\\
        $\textit{Norm}^\textit{flight}$ & The coefficient to normalize the value of the summation of flight shedding rates for each passenger.\\ 
        $\textit{Norm}^\textit{store}$ & The coefficient to normalize the value of the summation of storing shedding rates for each passenger.\\ 
        $M$ & Big positive value (here we suppose it takes 2)\\
        ~\\
        \multicolumn{2}{l}{\emph{Decision Variables}}\\
        $x_{ijb}$ & Binary variable, equals one if a passenger with the $(b-1)$ number of bag(s) is seated in a seat 
        in row $i$ and column $j$; equals zero otherwise\\
        $d_{ijb}^\textit{store}$ & The summation of shedding rates, that other passengers can cause when they are storing 
        bags, for a passenger who is seated in a seat in row $i$ and column $j$ having $(b-1)$ bag(s)\\
        $d_{ij}^\textit{flight}$ & The summation of shedding rates that the other passengers can cause for a passenger,
        who is seated in a seat in row $i$ and column $j$ during the flight\\
    \end{tabular}%
    \label{tab:param1}
\end{table}

\subsection{Transmission risk}
\label{sec:transmissionrisk}
We optimize the passenger boarding process and seat allocation, considering the boarding time and the risk associated with passenger interaction during movements in the aisle, storing luggage in the overhead compartment, and seating.  

Transmission risk can be defined by proximity to the index case and duration of contact time. Our approach is based on a transmission model~\citep{smieszek_epidemicsmodel_2009}, which defines the spread of SARS-CoV2 coronavirus as a function of (continuous) distance, using different distance measures~\citep{nagel_mobilityberlin_2020}. Here, the probability of a person $n$ being infected by a person $m$ is described by \eqref{eq:mechanistic_model}.
\begin{equation}
    P_{ n} = 1 - exp \left( -\theta \sum_m \sum_t \ \textit{SR}_{ m,t } \quad i_{ nm,t } \quad t_{ nm,t }  \right)
    \label{eq:mechanistic_model}
\end{equation}

defined by:

\begin{itemize}[leftmargin=1.2cm]
    \item[$ P_{ n }$]  Probability of person $n$ to receive an infectious dose. 
    Not ``infection probability'', which depends highly on the immune response of the affected person. 
    \item[$ \theta$]  Calibration factor for the specific disease.
    \item[$ \textit{SR}_{m,t}$] Shedding rate, the~amount of virus the person $m$ spreads during the timestep $t$.
    \item[$ i_{nm,t}$]  Intensity of the contact between $n$ and $m$  during the timestep $t$, which corresponds to their distance.
    \item[$ t_{ nm,t }$] Time person $n$ interacts with person $m$ at timestep $t$.
\end{itemize}

Considering this idea, we define the shedding rate \textit{SR} as a normalized bell-shaped function (\ref{eq:SR}) with $z \in (x, y)$ for both longitudinal and lateral dimensions, respectively. The parameters are $a$ (scaling factor), $b$ (slope of leading and falling edge), and $c$ (offset) to determine curve shape. 

\begin{equation}    
    \label{eq:SR}
    \text{SR}_{xy} = \prod_{z \in (x, y)} \left( 1+ \frac{|z - c_z|}{a_z} ^{2 b_z} \right) ^{-1}
\end{equation}

\textit{SR} was calibrated in a prior study \citep{schultz_evaluation_2020} based on the transmission events of an actual flight \citep{Olsen_Transmission_2003}. We have applied the corresponding parameter setting with $a_{x}=0.6$, $b_{x}=2.5$, $c_{x}=0.25$, $a_{y}=0.65$, $b_{y}=2.7$, and $c_{y}=0$. This causes the footprint in the y-direction (lateral to the direction of motion) to be smaller than in the x-direction (in the direction of motion). When passengers reach their seat row and start to store the hand luggage or enter the seat row, the direction of movement is changed by 90$^\circ$, heading to the aircraft window. Finally, the individual probability for virus transmission $P_n$ corresponds to $\Theta$, the specific intensity per timestep (\ref{eq:pn_theta}). 

\begin{equation}    
    P_n = \Theta \, \, \text{SR}_{xy} \, \, \alpha
    \label{eq:pn_theta}
\end{equation}

In accordance with \citep{schultz_evaluation_2020}, $\Theta$ is set to $\frac{1}{20}$, which means a passenger reaches the maximum probability of $P_n = 1$ after standing 20~s in closest distance in front of an infected passenger (SR$_{xy}=1$). The parameter $\alpha \in \{1,2\}$ is 1 and changed to 2 when the passenger stores the luggage or enters the seat row. This doubled shedding rate reflects the higher physical activities within a short distance to surrounding passengers. Since the probability $P_n$ is limited to 100\%, it is set to this value if the value determined by (\ref{eq:pn_theta}) is greater than 1.

This spatially continuous approach is applied to the agent-based simulation model to evaluate different boarding sequences (see Section~\ref{sec:agentSimulation}). For mixed-integer optimization, we consider corresponding discretized shedding rates for two distinctive situations: when a passenger is seated (flight shedding rates) and when the passenger is storing luggage (store shedding rates). 

\subsubsection{Flight shedding rates}
\begin{figure}[htb!]
    \centering
    \includegraphics[width=.4\textwidth]{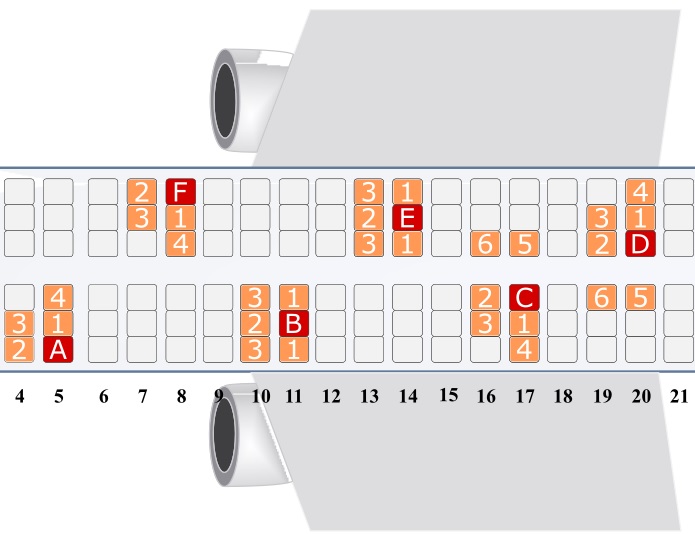}
    \caption{Types of passenger interactions (orange) in the aircraft cabin around the infected passengers (red) considering different seat positions: besides (\emph{SR}$^{\text{flight}}$ type 1 and type 4), in front (\emph{SR}$^{\text{flight}}$ type 2), diagonally in front (\emph{SR}$^{\text{flight}}$ type 3), and across the aisle (\emph{SR}$^{\text{flight}}$ type 5 and type 6)~\citep{Michael_Majid_TRC_2021}.}
    \label{fig:interactions}
\end{figure}

If an infected passenger seats to different columns, then the passengers around him/her could receive different amount of virus. For instance in Fig.~\ref{fig:interactions}, when a passenger seated in row $i=20$ and column D (aisle), we compute the shedding rate for the passenger that might seat in the same row ($i=20$ at column C (aisle, \emph{SR}$_{5}^{\text{flight}}$), E (middle, \emph{SR}$_{1}^{\text{flight}}$), and F (window, \emph{SR}$_{4}^{\text{flight}}$)) and previous row $i-1=19$ (column C (aisle, \emph{SR}$_{6}^{\text{flight}}$), D (aisle, \emph{SR}$_{2}^{\text{flight}}$), and E (middle, \emph{SR}$_{3}^{\text{flight}}$)).

Concerning the former situation, we followed the six fight shedding rate types provided by~\cite{Michael_Majid_TRC_2021}, considering an aisle/ seats width of 0.4m and a seat pitch of 0.8m. Thus, the shedding rate when a passenger is seated are:
\emph{SR}$_{1}^{\text{flight}} = 0.99987$, \emph{SR}$_{2}^{\text{flight}} = 0.9226$, \emph{SR}$_{3}^{\text{flight}} = 0.9126$, \emph{SR}$_{4}^{\text{flight}} = \ $\emph{SR}$_{5}^{\text{flight}} = 0.6833$, and \emph{SR}$_{6}^{\text{flight}} = 0.6315$.

\subsubsection{Store shedding rates}
\begin{figure}[htb!]
    \centering
    \includegraphics[width=.4\textwidth]{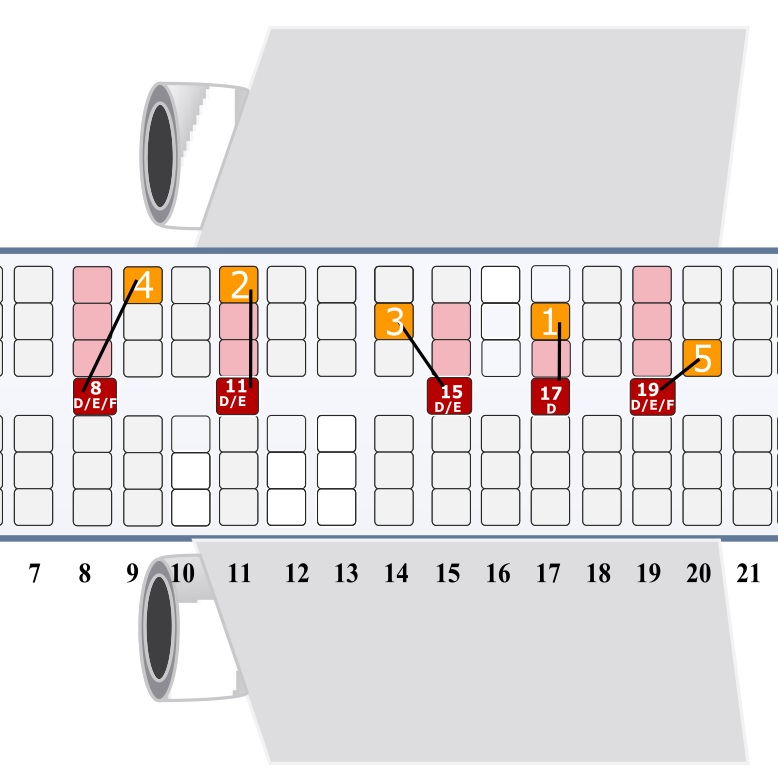}
    \caption{Types of passenger interactions (orange) in the aircraft cabin for the  passengers (9F, 11F, 14E, 17E and 20D) considering different distances of the infected  passengers who are storing their bag(s) in the aisle (red).}
    \label{fig:storesr}
\end{figure}

We define the second type of shedding rates for the passengers when they are storing their bags. The first shedding rate was defined just based on the distances between two passengers. The second type defines not only the distance between a passenger who keeps his luggage and another passenger who sits earlier (see Fig.~\ref{fig:storesr}), but also the number of pieces of luggage that are already in the overhead bins. The store shedding rate when a passenger is storing its bags are: \emph{SR}$_{1}^{\text{store}} = 0.6833$, \emph{SR}$_{2}^{\text{store}} = 0.1951$, \emph{SR}$_{3}^{\text{store}} = 0.6315$,  \emph{SR}$_{4}^{\text{store}} = 0.1803$, and \emph{SR}$_{5}^{\text{store}} = 0.9126$. 
The store shedding rate types 3, 4, and 5 relate to infected passengers of the previous or next row that could create target passengers (14E, 9F, and 20D). For example, for passenger 9F, we consider the infected passenger (coded red) in the aisle, who might seat in 8D, 8E, or 8F. However, when s/he is storing the bags, the distance to the orange-coded passenger (9F), is the same and we consider type 4 for this circumstance. Also, the store shedding rate types 1 and 2 are created where a passenger (orange-coded) is infected by another passenger (red-coded) who seats in the same row but after orange-coded passenger, when red-coded passenger stores its bags (see Fig.~\ref{fig:storesr}). We show the type of interactions with a number in orange-coded seats here. For example, infected passenger 17D, shown as a red-coded passenger, creates interaction type 1 for passenger 17E (orange-coded) when (s)he is storing bags. In addition, the seats of infected passengers in the aisle are shown in Fig.~\ref{fig:storesr} with pink color cells. 

Here, we face three situations in general. In the first case, we calculate the second shedding rate for a passenger who seats in the window seat (i.e., column A or F). In the second case, the passenger seats on the middle seat (i.e., column B or E), and finally who seats on the aisle seat (i.e., column C or D). As the strategy that we used to board the passengers is outside-in, we can define the storing shedding rate. 

\begin{figure*}[htb!]
 \centering
   \includegraphics[width=.7\textwidth]{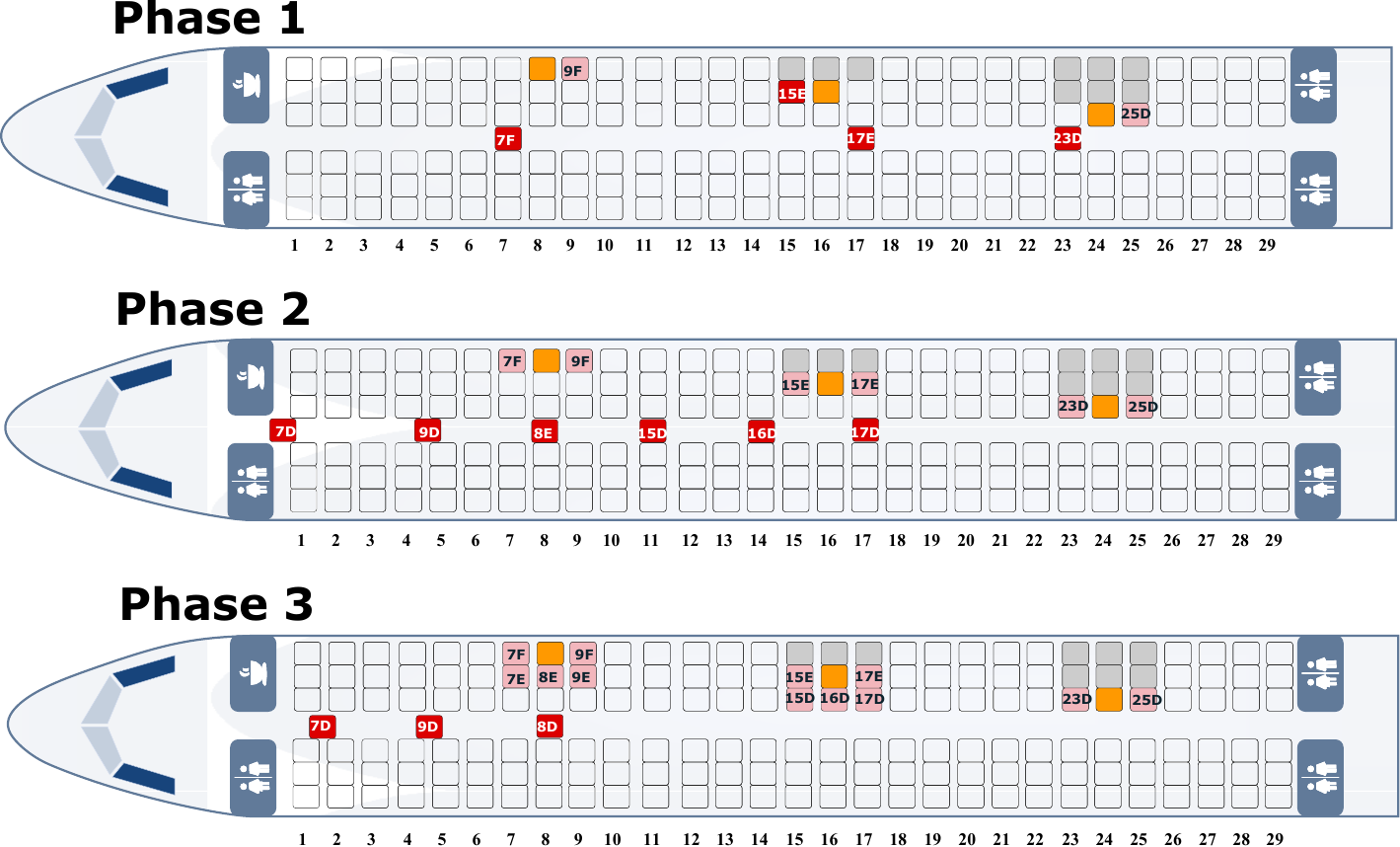}
 \caption{The process of boarding infected passengers (red-coded)  near target passengers (orange-coded) in three phases where they are storing their bags}
 \label{fig:SR2}
\end{figure*}

In an exemplary boarding process, we show how to calculate the sum of storing shedding rates. Fig.~\ref{fig:SR2} indicates the different interactions that happen for target passengers who seat at the window (8F), middle (16E), and aisle (24D) seats. We just highlight the right side of the aisle because for the left side, the process is the same. First for the passenger who seats on 8F (window seat). To calculate the storing shedding rate for this passenger, we use (\ref{eq:threephase}), which includes the three elements associated with the outside-in boarding.

\begin{equation}
    d_{8,6,b}^\textit{store} = \textit{SR}_\text{Phase 1} + \textit{SR}_\text{Phase 2}+\textit{SR}_\text{Phase 3} 
    \label{eq:threephase}
\end{equation}   

The first element is defined based on the first phase of boarding passengers, i.e., for passengers sitting at the window (column F), near the orange passenger (seat 8F). The second element is used to calculate the shedding rates of passengers sitting in middle seats (column E), and the last element includes the shedding rates of passengers sitting in aisle seats (column D).

\paragraph{\textbf{Phase 1}}
  In (\ref{eq:p1}) we calculate the corresponding shedding rate for window-seated passengers.
 \begin{equation}
    \textit{SR}_\text{Phase 1}=0.25\sum_{b'=2}^{\mathcal{B}} (b+b') \, \text{\textit{SR}}_{4}^\textit{store} \, \{ {x_{7,6,b'}+x_{9,6,b'}} \}
    \label{eq:p1}
\end{equation}   
 
We use the coefficient of 0.25 because, we consider 0.5 probability for those two infected passengers, 7F and 9F, because they could be seated sooner or later than the orange-coded passenger (see  Fig.~\ref{fig:SR2}). Also, it multiplies another 0.5, because they seat in the previous or next row. In this formulation, we consider $b+b'$ which is the sum of bags of the target passenger (8F) and infected passengers (7F and 9F). The number of bags of them is important and when the number of bags is greater therefore the risk indicator will be increased. Therefore, we define this coefficient in the formulation. Finally, we have two binary decision variables for passengers 7F and 9F with $(b'-1)$ bags. \emph{SR}$_{4}^\textit{store}$ is the shedding rate based on the distance between the target passenger and infected passengers when they are storing their bags in the aisle. In Fig.~\ref{fig:SR2}, target passengers are shown with orange cells, infected passengers who are seated with pink cells, and infected passengers who are in the aisle and trying to store their bags with red cells. 
 
To calculate the correct value of storing shedding rate for the target passenger with seat 8F in Fig.~\ref{fig:SR2}, we need to calculate the shedding rates that passengers who seat in the middle and aisle seats of rows 7, 8, and 9 in the next phase of boarding in the sum of store shedding rates.

If we want to calculate the sum of store shedding rate for another target passenger in the middle seat (16E), we have the same formulation, the only difference is instead of \emph{SR}$_{4}^\textit{store}$, we use \emph{SR}$_{3}^\textit{store}$ because for example, when infected passenger 15E or 17E is storing its bags, the distance between the infected passengers with the target passengers are different. Obviously, \emph{SR}$_{3}^\textit{store}$ is greater than \emph{SR}$_{4}^\textit{store}$. Also, we have like this situation for another target passenger who seats in 24D. In other words, the passenger with seat 23D leads to storing shedding rate \emph{SR}$_{5}^\textit{store}$ for that passenger. Also, \emph{SR}$_{5}^\textit{store}$ is greater than \emph{SR}$_{3}^\textit{store}$ because of the distance. We can see these concepts in Fig.~\ref{fig:storesr} as well when the distance of target passenger 9F to infected passenger 8D/ 8E/ 8F is greater than the distance of target passenger 14E to infected passenger 15D/ 15E.

\paragraph{\textbf{Phase 2}}
In the second phase of boarding, passengers located at middle seats store their luggage. Thus, we need to extend (\ref{eq:p1}) and add new  elements (\ref{eq:p2}).
 \begin{equation}
    \begin{aligned}
      \textit{SR}_\text{Phase 2} = & 0.5\sum_{b'=2}^{\mathcal{B}} (b+b') \  \text{\textit{SR}}_{4}^\textit{store}\{x_{7,5,b'}+x_{9,5,b'}\}\\
      &+ \sum_{b'=2}^{\mathcal{B}} (b+b') \  \text{\textit{SR}}_{2}^\textit{store}\{x_{8,5,b'}\}
      \label{eq:p2}
    \end{aligned}
\end{equation}
    
In the right hand of this equation, we calculate the sum of shedding rates that passengers of middle seats can lead. This part is defined based on passengers with seats 7E and 9E (with shedding rates type 4) and the last part (with shedding rates type 2), is defined based on the location of the passenger with seat 8E in Fig.~\ref{fig:SR2}. The order of these three passengers is not important, as they can all transfer viruses to the destination passenger in seat 8F if they have baggage. The main reason is that we use the outside-in strategy for the boarding phase. Therefore, they seat after the target passenger with seat 8F. In addition, for that passenger, we need to calculate the shedding rates that passengers of aisle seats can lead. We explain it in phase 3. 

Also, we have a similar situation for the target passenger at seat 16E and three passengers with seats 15D, 16D, and 17D. We consider shedding rate type 3, \emph{SR}$_{3}^\textit{store}$, for passengers with seats 15D and 17D and shedding rate type 1, \emph{SR}$_{1}^\textit{store}$, for passenger with seat 16D that spread for that target passenger. Like the last part, we can see this concept in Fig.~\ref{fig:storesr} as well. These two types of store shedding rates are shown with the number of orange-coded passengers there.

\paragraph{\textbf{Phase 3}} 
The last element considers aisle-seated passengers (\ref{eq:p3}). For the proposed target passenger (seat 8F), these passengers are in seats 7D, 8D, and 9D. 

\begin{equation}
    \begin{aligned}
      \textit{SR}_\text{Phase 3}= &0.5\sum_{b'=2}^{\mathcal{B}} (b+b') \  \textit{SR}_{4}^\textit{store}\{x_{7,4,b'}+x_{9,4,b'}\}\\&+ \sum_{b'=2}^{\mathcal{B}} (b+b')\ \textit{SR}_{2}^\textit{store}\{x_{8,4,b'}\}
      \label{eq:p3}
    \end{aligned}
\end{equation}

Also, we use the same shedding rate, \emph{SR}$_{4}^\textit{store}$, for passengers 7D, 7E, 7F, 9D, 9E, 9F for target passenger 8F because where they are storing their bags, the distances between each on them and passenger 8F is the same. 

\subsection{Constraints}
To understand the equations, we explain below how we model the most important constraints, first with respect to the shedding rates during luggage storage. The notations and definitions of parameters and decision variables are defined as follows.

The proposed mixed-integer linear programming model for the problem is introduced with (\ref{eq:5})-(\ref{eq:33}). 
In the objective function, we minimize the summation of flight and store shedding rates of all passengers in~(\ref{eq:objFunc}). If there is an empty seat, we do not calculate the shedding rates here, based on the constraint (\ref{eq:5}). Constraint (\ref{eq:6}) guarantee that each passenger only has one choice between 0, 1, or 2 bags. The number of passengers who have $(b-1)$ bags is determined by constraint (\ref{eq:7}). 

Equations (\ref{eq:8})-(\ref{eq:17}) were already defined and successfully implemented in \citep{Michael_Majid_TRC_2021}. We follow this approach and update these equations according to our requirements. We add a coefficient, $\textit{Norm}^\textit{flight} =\frac{1}{4.8209}$, which helps us to create the normalized value of the summation of flight shedding rates for each passenger. With (\ref{eq:8})-(\ref{eq:12}), we calculate the sum of flight shedding rates for passengers who seat in the first row. The row count starts with $i$=1 (no row number 0). In fact, we restricted the number of rows because passengers sitting in the previous row are important for calculating the flight shedding rate. The sum of flight shedding rates of passengers who seat from the second row to the last row is calculated based on (\ref{eq:13})-(\ref{eq:17}).

The constraints (\ref{eq:18}) and (\ref{eq:19}) guarantee to compute the storing shedding rates for passengers with window seats (i.e., column $A$, $F$, or $j$=1, 6). For example, for the passenger who seats on a window seat (8F) we calculate the storing shedding rates in the last part and indicate in Fig.~\ref{fig:SR2} with details. The storing shedding rates for passengers with middle seats (i.e., column $B$, $E$, or $j$=2, 5) are calculated by constraints (\ref{eq:20}) and (\ref{eq:21}) and calculated for passengers with aisle seats by constraint (\ref{eq:22}). Similarly, we use a coefficient, $\textit{Norm}^\textit{flight}=\frac{1}{9.7833}$, which helps us to create the normalized value of the summation of storing shedding rates for each passenger. This value, like flight shedding rate, is the maximum shedding rate that can occur for a passenger when the passenger and the other passengers who seat near that passenger, have 2 bags.

To calculate the sum of storing shedding rates for passengers of window seats, middle seats and aisle seats, constraints (\ref{eq:23}-\ref{eq:24}), constraints (\ref{eq:25}-\ref{eq:26}) and constraint (\ref{eq:27}) are considered, respectively. Similarly, constraints (\ref{eq:28}-\ref{eq:29}), constraints (\ref{eq:30}-\ref{eq:31}) and constraint (\ref{eq:32}) are defined to compute the sum of storing shedding rates for passengers of window seats, middle seats and aisle seats in the last row of the cabin. In addition, we do not need to consider the shedding rates for the next row because they are seated on the last row. Finally, the range of the decision variable is defined in constraint~(\ref{eq:33}).

\begin{figure*}
    \scriptsize
\begin{align}
     \hspace{10pt}  min\hspace{4pt} \sum_{i=1}^{\mathcal{I}}\sum_{j=1}^{\mathcal{J}}
    d_{ij}^\textit{flight}+\sum_{i=1}^{\mathcal{I}}\sum_{j=1}^{\mathcal{J}}\sum_{b=1}^{\mathcal{B}}d_{ijb}^\textit{store}\label{eq:objFunc}
\end{align}

\begingroup
    \allowdisplaybreaks
    \scriptsize
    \label{eq:model}
    \begin{align}
        &  d_{ij}^\textit{flight}+\sum_{b=1}^{\mathcal{B}}d_{ijb}^\textit{store} \le 2 \sum_{b=1}^{\mathcal{B}} x_{ijb}  \hspace{10cm} \forall \quad i,j \label{eq:5}\\
        &  \sum_{b=1}^{\mathcal{B}} x_{ijb} \le 1 \hspace{12.1cm} \forall \quad i,j \label{eq:6}\\
        & \sum_{i=1}^{\mathcal{I}}\sum_{j=1}^{\mathcal{J}} x_{ijb} = \textit{Bag}_{b} \hspace{11.2cm}\forall \quad b \label{eq:7}\\
        & M \left( \sum_{b=1}^{\mathcal{B}}x_{ijb}-1 \right) +\textit{Norm}^\textit{flight}\Bigg[ \sum_{b'=1}^{\mathcal{B}} \{ \textit{SR}_{1}^\textit{flight}x_{i(j+1)b'} +  \textit{SR}_{4}^\textit{flight}x_{i(j+2)b'} \} \Bigg]\le d_{ij}^\textit{flight} \hspace{4.9cm}\forall \quad i=1,j=1 \label{eq:8}\\
        & M \left( \sum_{b=1}^{\mathcal{B}}x_{ijb}-1 \right) +\textit{Norm}^\textit{flight}\Bigg[  \sum_{b'=1}^{\mathcal{B}} \textit{SR}_{1}^\textit{flight} \{ x_{i(j-1)b'} + x_{i(j+1)b'} \} \Bigg]\le d_{ij}^\textit{flight}  \hspace{5.7cm} \forall \quad i=1,j=2,5 \label{eq:9}\\
        & M \left( \sum_{b=1}^{\mathcal{B}}x_{ijb}-1 \right) +\textit{Norm}^\textit{flight}\Bigg[  \sum_{b'=1}^{\mathcal{B}}  \{ \textit{SR}_{4}^\textit{flight}x_{i(j-2)b'} + \textit{SR}_{1}^\textit{flight}x_{i(j-1)b'} + \textit{SR}_{5}^\textit{flight}x_{i(j+1)b'} \}\Bigg] \le d_{ij}^\textit{flight} \hspace{2.8cm}\forall \quad i=1,j=3 \label{eq:10}\\
        & M \left( \sum_{b=1}^{\mathcal{B}}x_{ijb}-1 \right) +\textit{Norm}^\textit{flight}\Bigg[  \sum_{b'=1}^{\mathcal{B}} \{ \textit{SR}_{5}^\textit{flight}x_{i(j-1)b'} + \textit{SR}_{1}^\textit{flight}x_{i(j+1)b'} + \textit{SR}_{4}^\textit{flight}x_{i(j+2)b'} \} \Bigg]\le d_{ij}^\textit{flight}  \hspace{2.8cm} \forall \quad i=1,j=4 \label{eq:11}\\
        & M \left( \sum_{b=1}^{\mathcal{B}}x_{ijb}-1 \right) +\textit{Norm}^\textit{flight}\Bigg[  \sum_{b'=1}^{\mathcal{B}} \{ \textit{SR}_{4}^\textit{flight}x_{i(j-2)b'} + \textit{SR}_{1}^\textit{flight}x_{i(j-1)b'}\} \Bigg] \le d_{ij}^\textit{flight}  \hspace{4.8cm} \forall \quad i=1,j=6 \label{eq:12}\\
        & M \left( \sum_{b=1}^{\mathcal{B}}x_{ijb}-1 \right) +\textit{Norm}^\textit{flight}\Bigg[  \sum_{b'=1}^{\mathcal{B}} \{ \textit{SR}_{2}^\textit{flight}x_{(i-1)jb'} +  \textit{SR}_{3}^\textit{flight}x_{(i-1)(j+1)b'} + \textit{SR}_{1}^\textit{flight}x_{i(j+1)b'} + \textit{SR}_{4}^\textit{flight}x_{i(j+2)b'} \} \Bigg] \le d_{ij}^\textit{flight}  \hspace{0.4cm} \forall \quad i\ge 2,j=1 \label{eq:13}\\
        & M \left( \sum_{b=1}^{\mathcal{B}}x_{ijb}-1 \right) +\textit{Norm}^\textit{flight}\Bigg[  \sum_{b'=1}^{\mathcal{B}} \{ \textit{SR}_{3}^\textit{flight}x_{(i-1)(j-1)b'} + \textit{SR}_{2}^\textit{flight}x_{(i-1)jb'}+ \textit{SR}_{3}^\textit{flight}x_{(i-1)(j+1)b'} + \textit{SR}_{1}^\textit{flight}x_{i(j-1)b'}+ \textit{SR}_{1}^\textit{flight}x_{i(j+1)b'} \} \Bigg]\le d_{ij}^\textit{flight} \nonumber \\
        & \hspace{13.4cm} \forall \quad i\ge 2,j=2,5 \label{eq:14}\\
        & M \left( \sum_{b=1}^{\mathcal{B}}x_{ijb}-1 \right) +\textit{Norm}^\textit{flight}\Bigg[  \sum_{b'=1}^{\mathcal{B}} \{ \textit{SR}_{3}^\textit{flight}x_{(i-1)(j-1)b'} + \textit{SR}_{2}^\textit{flight}x_{(i-1)jb'}+ \textit{SR}_{6}^\textit{flight}x_{(i-1)(j+1)b'}
        + \textit{SR}_{4}^\textit{flight}x_{i(j-2)b'}+ \textit{SR}_{1}^\textit{flight}x_{i(j-1)b'} \nonumber \\
        & \hspace{3cm}+ \textit{SR}_{5}^\textit{flight}x_{i(j+1)b'}\} \Bigg]\le d_{ij}^\textit{flight} \hspace{7.2cm}\forall \quad i\ge 2,j=3 \label{eq:15}\\
        & M \left( \sum_{b=1}^{\mathcal{B}}x_{ijb}-1 \right) +\textit{Norm}^\textit{flight}\Bigg[  \sum_{b'=1}^{\mathcal{B}} \{\textit{SR}_{6}^\textit{flight}x_{(i-1)(j-1)b'} + \textit{SR}_{2}^\textit{flight}x_{(i-1)jb'}+ \textit{SR}_{3}^\textit{flight}x_{(i-1)(j+1)b'} + \textit{SR}_{5}^\textit{flight}x_{i(j-1)b'}+ \textit{SR}_{1}^\textit{flight}x_{i(j+1)b'} \nonumber\\
        & \hspace{3cm}+ \textit{SR}_{4}^\textit{flight}x_{i(j+2)b'} \}  \Bigg]\le d_{ij}^\textit{flight} \hspace{7.2cm} \forall \quad i\ge 2,j=4 \label{eq:16}\\
        & M \left( \sum_{b=1}^{\mathcal{B}}x_{ijb}-1 \right) +\textit{Norm}^\textit{flight}\Bigg[  \sum_{b'=1}^{\mathcal{B}} \{\textit{SR}_{3}^\textit{flight}x_{(i-1)(j-1)b'} + \textit{SR}_{2}^\textit{flight}x_{(i-1)jb'}+\textit{SR}_{4}^\textit{flight}x_{i(j-2)b'} + \textit{SR}_{1}^\textit{flight}x_{i(j-1)b'}\} \Bigg]\le d_{ij}^\textit{flight} \hspace{0.5cm} \forall \quad i\ge 2,j=6 \label{eq:17}\\
        & M \left( \sum_{b'=1}^{\mathcal{B}}x_{ijb'}-1 \right) +\textit{Norm}^\textit{store}\Bigg[ 0.25\sum_{b'=2}^{\mathcal{B}} (b+b') \textit{SR}_{4}^\textit{store}\{{x_{(i-1)jb')}+x_{(i+1)jb'}}\} +0.5\sum_{b'=2}^{\mathcal{B}} (b+b') \textit{SR}_{4}^\textit{store}\{x_{(i-1)(j-1)b'}+x_{(i+1)(j-1)b'}+x_{(i-1)(j-2)b'}+x_{(i+1)(j-2)b'}\}
        \nonumber \\
        &\hspace{3cm} +\sum_{b'=2}^{\mathcal{B}} (b+b') \textit{SR}_{2}^\textit{store}\{x_{(i)(j-1)b'}+x_{(i)(j-2)b'}\} \Bigg] \le d_{ijb}^\textit{store} \hspace{4.7cm}  \forall \quad  b,2\le i\le 28, j=6 \label{eq:18}\\
        & M \left( \sum_{b'=1}^{\mathcal{B}}x_{ijb'}-1 \right) +\textit{Norm}^\textit{store}\Bigg[ 0.25\sum_{b'=2}^{\mathcal{B}} (b+b') \textit{SR}_{4}^\textit{store}\{{x_{(i-1)jb')}+x_{(i+1)jb'}}\} +0.5\sum_{b'=2}^{\mathcal{B}} (b+b') \textit{SR}_{4}^\textit{store}\{x_{(i-1)(j+1)b'}+x_{(i+1)(j+1)b'}+x_{(i-1)(j+2)b'}+x_{(i+1)(j+2)b'}\}
        \nonumber \\
        &\hspace{3cm}+\sum_{b'=2}^{\mathcal{B}} (b+b') \textit{SR}_{2}^\textit{store}\{x_{(i)(j+1)b'}+x_{(i)(j+2)b'}\} \Bigg] \le  d_{ijb}^\textit{store}
         \hspace{4.7cm}  \forall \quad  b,2\le i\le 28, j=1 \label{eq:19}\\
        & M \left( \sum_{b'=1}^{\mathcal{B}}x_{ijb'}-1 \right) +\textit{Norm}^\textit{store}\Bigg[ 0.25\sum_{b'=2}^{\mathcal{B}} (b+b') \textit{SR}_{3}^\textit{store}\{{x_{(i-1)jb'}+x_{(i+1)jb'}}\}+\sum_{b'=2}^{\mathcal{B}} (b+b') \textit{SR}_{1}^\textit{store}\{x_{(i)(j-1)b'}\}
        \nonumber \\
        &\hspace{3cm}+0.5\sum_{b'=2}^{\mathcal{B}} (b+b') \textit{SR}_{3}^\textit{store}\{x_{(i-1)(j-1)b'}+x_{(i+1)(j-1)b'}\}
             \Bigg]   \le  d_{ijb}^\textit{store} \hspace{3.9 cm} \forall \quad  b,2\le i\le 28, j=5 \label{eq:20}
        \end{align}
\endgroup
\end{figure*}

\begin{figure*}
\begingroup
    \scriptsize
    \allowdisplaybreaks
    \begin{align}         
        & M \left( \sum_{b'=1}^{\mathcal{B}}x_{ijb'}-1 \right) +\textit{Norm}^\textit{store}\Bigg[ 0.25\sum_{b'=2}^{\mathcal{B}} (b+b') \textit{SR}_{3}^\textit{store}\{{x_{(i-1)jb'}+x_{(i+1)jb'}}\}+\sum_{b'=2}^{\mathcal{B}} (b+b') \textit{SR}_{1}^\textit{store}\{x_{(i)(j+1)b'}\}
        \nonumber \\
        &\hspace{3cm}+0.5\sum_{b'=2}^{\mathcal{B}} (b+b') \textit{SR}_{3}^\textit{store}\{x_{(i-1)(j+1)b'}+x_{(i+1)(j+1)b'}\}
               \Bigg]   \le  d_{ijb}^\textit{store}
        \hspace{4 cm} \forall \quad  b,2\le i\le 28, j=2 \label{eq:21}\\
        & M \left( \sum_{b'=1}^{\mathcal{B}}x_{ijb'}-1 \right) +\textit{Norm}^\textit{store}\Bigg[ 0.25\sum_{b'=2}^{\mathcal{B}} (b+b') \textit{SR}_{5}^\textit{store}\{{x_{(i-1)jb'}+x_{(i+1)jb'}}\} \Bigg]   \le  d_{ijb}^\textit{store}
        \hspace{5 cm}\forall \quad  b,2\le i\le 28, j=3,4 \label{eq:22}\\
        & M \left( \sum_{b'=1}^{\mathcal{B}}x_{ijb'}-1 \right) +\textit{Norm}^\textit{store}\Bigg[ 0.25\sum_{b'=2}^{\mathcal{B}} (b+b') \textit{SR}_{4}^\textit{store}\{x_{(i+1)jb'}\} +0.5\sum_{b'=2}^{\mathcal{B}} (b+b') \textit{SR}_{4}^\textit{store}\{x_{(i+1)(j-1)b'}+x_{(i+1)(j-2)b'}\}
        \nonumber \\
        &\hspace{3cm}+\sum_{b'=2}^{\mathcal{B}} (b+b') \textit{SR}_{2}^\textit{store}\{x_{(i)(j-1)b'}+x_{(i)(j-2)b'}\}
               \Bigg]   \le  d_{ijb}^\textit{store} \hspace{5 cm} \forall \quad  b,\quad i=1, j=6 \label{eq:23}\\
        & M \left( \sum_{b'=1}^{\mathcal{B}}x_{ijb'}-1 \right) +\textit{Norm}^\textit{store}\Bigg[ 0.25\sum_{b'=2}^{\mathcal{B}} (b+b') \textit{SR}_{4}^\textit{store}\{x_{(i+1)jb'}\} +0.5\sum_{b'=2}^{\mathcal{B}} (b+b') \textit{SR}_{4}^\textit{store}\{x_{(i+1)(j+1)b'}+x_{(i+1)(j+2)b'}\}
        \nonumber \\
        &\hspace{3cm}+\sum_{b'=2}^{\mathcal{B}} (b+b') \textit{SR}_{2}^\textit{store}\{x_{(i)(j+1)b'}+x_{(i)(j+2)b'}\}
              \Bigg]   \le  d_{ijb}^\textit{store} \hspace{5 cm}\forall \quad  b,\quad i=1, j=1 \label{eq:24}\\
        & M \left( \sum_{b'=1}^{\mathcal{B}}x_{ijb'}-1 \right) +\textit{Norm}^\textit{store}\Bigg[ 0.25\sum_{b'=2}^{\mathcal{B}} (b+b') \textit{SR}_{3}^\textit{store}\{x_{(i+1)jb'}\}+0.5\sum_{b'=2}^{\mathcal{B}} (b+b') \textit{SR}_{3}^\textit{store}\{x_{(i+1)(j-1)b'}\}
        \nonumber \\
        &\hspace{3cm} +\sum_{b'=2}^{\mathcal{B}} (b+b') \textit{SR}_{1}^\textit{store}\{x_{(i)(j-1)b'}\} \Bigg]   \le  d_{ijb}^\textit{store}
         \hspace{6.4 cm}\forall \quad  b,\quad i=1, j=5 \label{eq:25}\\
        & M \left( \sum_{b'=1}^{\mathcal{B}}x_{ijb'}-1 \right) +\textit{Norm}^\textit{store}\Bigg[ 0.25\sum_{b'=2}^{\mathcal{B}} (b+b') \textit{SR}_{3}^\textit{store}\{x_{(i+1)jb'}\}+0.5\sum_{b'=2}^{\mathcal{B}} (b+b') \textit{SR}_{3}^\textit{store}\{x_{(i+1)(j+1)b'}\}
        \nonumber \\
        &\hspace{3cm}+\sum_{b'=2}^{\mathcal{B}} (b+b') \textit{SR}_{1}^\textit{store}\{x_{(i)(j+1)b'}\}
               \Bigg]   \le  d_{ijb}^\textit{store} \hspace{6.4 cm}\forall \quad  b,\quad i=1, j=2 \label{eq:26}\\
        & M \left( \sum_{b'=1}^{\mathcal{B}}x_{ijb'}-1 \right) +\textit{Norm}^\textit{store}\Bigg[ 0.25\sum_{b'=2}^{\mathcal{B}} (b+b') \textit{SR}_{5}^\textit{store}\{x_{(i+1)jb'}\}             \Bigg]   \le  d_{ijb}^\textit{store}
        \quad \hspace{6.1 cm}\forall \quad  b,\quad i=1, j=3,4 \label{eq:27}\\
      & M \left( \sum_{b'=1}^{\mathcal{B}}x_{ijb'}-1 \right) +\textit{Norm}^\textit{store}\Bigg[ 0.25\sum_{b'=2}^{\mathcal{B}} (b+b') \textit{SR}_{4}^\textit{store}\{x_{(i-1)jb'}\}  +0.5\sum_{b'=2}^{\mathcal{B}} (b+b') \textit{SR}_{4}^\textit{store}\{x_{(i-1)(j-1)b'}+x_{(i-1)(j-2)b'}\}
        \nonumber \\
        &\hspace{3cm} +\sum_{b'=2}^{\mathcal{B}} (b+b') \textit{SR}_{2}^\textit{store}\{x_{(i)(j-1)b'}+x_{(i)(j-2)b'}\}
              \Bigg]   \le  d_{ijb}^\textit{store} \hspace{5.2 cm}\forall \quad  b,\quad i=29, j=6 \label{eq:28}\\
        & M \left( \sum_{b'=1}^{\mathcal{B}}x_{ijb'}-1 \right) +\textit{Norm}^\textit{store}\Bigg[ 0.25\sum_{b'=2}^{\mathcal{B}} (b+b') \textit{SR}_{4}^\textit{store}\{x_{(i-1)jb'}\} +0.5\sum_{b'=2}^{\mathcal{B}} (b+b') \textit{SR}_{4}^\textit{store}\{x_{(i-1)(j+1)b'}+x_{(i-1)(j+2)b'}\}
        \nonumber \\
        &\hspace{3cm}+\sum_{b'=2}^{\mathcal{B}} (b+b') \textit{SR}_{2}^\textit{store}\{x_{(i)(j+1)b'}+x_{(i)(j+2)b'}\} 
        \Bigg]   \le  d_{ijb}^\textit{store}
        \hspace{5.2 cm}\forall \quad  b,\quad i=29, j=1 \label{eq:29}\\
        & M \left( \sum_{b'=1}^{\mathcal{B}}x_{ijb'}-1 \right) +\textit{Norm}^\textit{store}\Bigg[ 0.25\sum_{b'=2}^{\mathcal{B}} (b+b') \textit{SR}_{3}^\textit{store}\{x_{(i-1)jb'}\}+0.5\sum_{b'=2}^{\mathcal{B}} (b+b') \textit{SR}_{3}^\textit{store}\{x_{(i-1)(j-1)b'}\}
        \nonumber \\
        &\hspace{3cm}+\sum_{b'=2}^{\mathcal{B}} (b+b') \textit{SR}_{1}^\textit{store}\{x_{(i)(j-1)b'}\}
              \Bigg]   \le  d_{ijb}^\textit{store}
         \hspace{6.6 cm}\forall \quad  b,\quad i=29, j=5 \label{eq:30}\\
        & M \left( \sum_{b'=1}^{\mathcal{B}}x_{ijb'}-1 \right) +\textit{Norm}^\textit{store}\Bigg[ 0.25\sum_{b'=2}^{\mathcal{B}} (b+b') \textit{SR}_{3}^\textit{store}\{x_{(i-1)jb'}\}+0.5\sum_{b'=2}^{\mathcal{B}} (b+b') \textit{SR}_{3}^\textit{store}\{x_{(i-1)(j+1)b'}\}
        \nonumber \\
        &\hspace{3cm}+\sum_{b'=2}^{\mathcal{B}} (b+b') \textit{SR}_{1}^\textit{store}\{x_{(i)(j+1)b'}\}
              \Bigg]   \le  d_{ijb}^\textit{store}
         \hspace{6.6 cm}\forall \quad  b,\quad i=29, j=2 \label{eq:31}\\
        & M \left( \sum_{b'=1}^{\mathcal{B}}x_{ijb'}-1 \right) +\textit{Norm}^\textit{store}\Bigg[ 0.25\sum_{b'=2}^{\mathcal{B}} (b+b') \textit{SR}_{5}^\textit{store}\{x_{(i-1)jb'}\}  \Bigg]   \le  d_{ijb}^\textit{store} 
        \quad \hspace{6.3 cm}\forall \quad  b,\quad i=29, j=3,4 \label{eq:32}\\
        & x_{ijb}\in \{0,1\}, \quad d_{ij}^\textit{flight},d_{ijb}^\textit{store}\ge 0    \quad\quad  \hspace{10.2 cm}\forall \quad i,j,b \label{eq:33}
    \end{align}
\endgroup
\end{figure*}
\normalsize

To calculate the store shedding rates, we had two options; the first option is the definition of a couple of constraints which include distance and bags of two groups of passengers. The second option is the proposed model in this study. We decided not to use the first option because in valuing this kind of parameter, we need to calculate too many combinations and more complexity should be considered because the formulation includes two arrays of three numbers. Each number shows the number of bags. Suppose that on the right side of the cabin in row $i$ we have (b, b', b'') which are the number of bags of passengers who sit in columns D, E, and F in the row $i$. Therefore, we have 3*3*3=27 combinations for the right side of a seat row. Then, if we just accept just a gap between the number of rows of one, therefore we have (number of rows-1=28)*27*27=20412 combinations for the right side of the cabin. Also, we should consider for the left side to find out all acceptable combinations and value them. Therefore, we decided to use the formulation that includes x and y of two passengers and multiply by the sum of bags of the passengers.

\section{Results of the proposed approach}
\label{sec:optresults}

\begin{figure}[htb!]
    \centering
    \includegraphics[width=.45\textwidth]{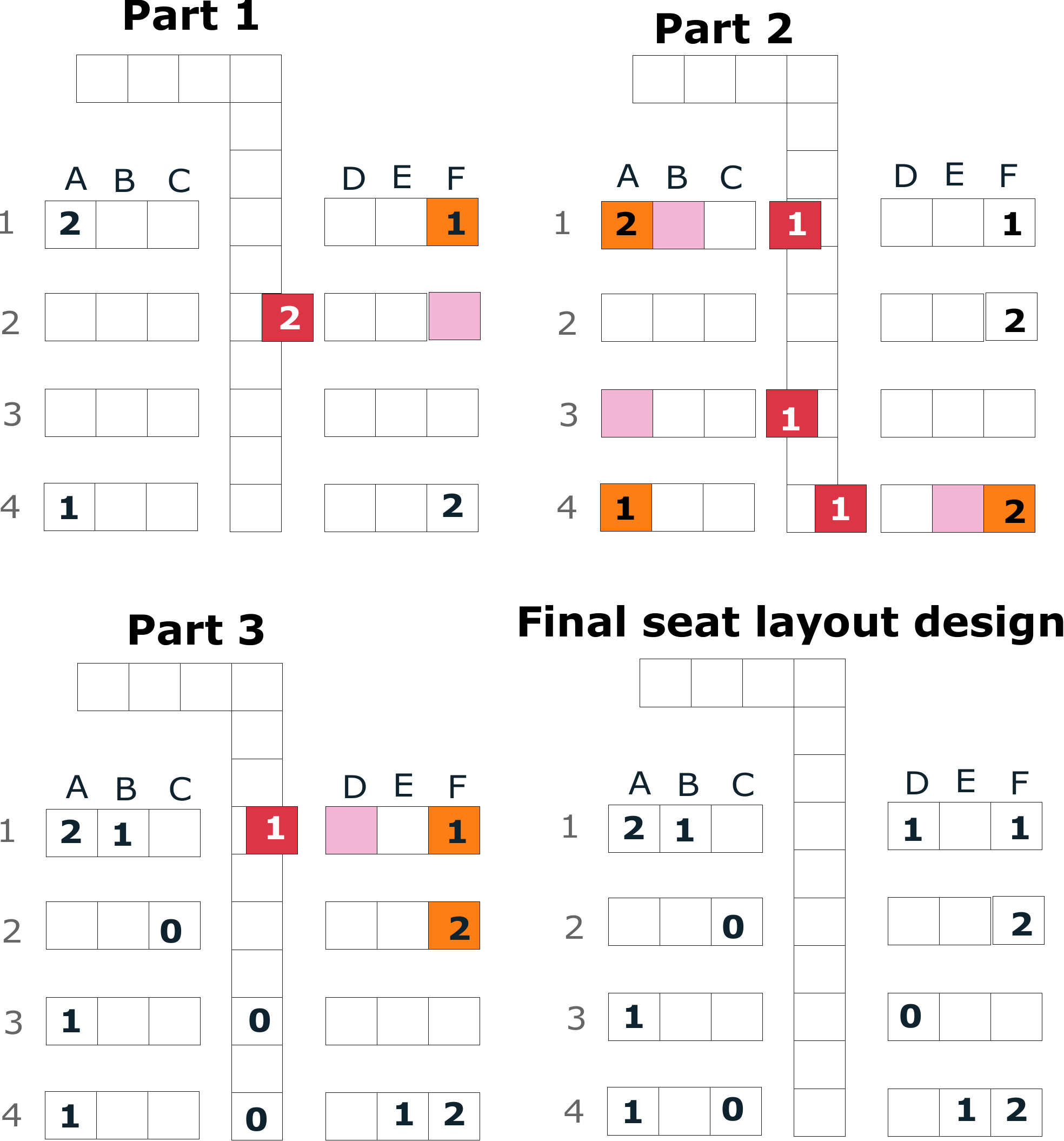}
    \caption{The optimal solution of small-size problem (numbers show the number of bags)}
    \label{fig:small4rows}
\end{figure}

Several small to medium-sized problems are designed to validate the model. These problems are solved with solver Gurobi in Python and the optimal solutions are determined. Fig.~\ref{fig:small4rows} shows the result of the smallest problem solved, where the number of rows is 4 and a seat load of 50\% is assumed. Furthermore, three passengers without luggage, six passengers with one piece of luggage, and three passengers with two pieces of luggage are considered. The run time for this use case is less than 1 minute and the value of the objective function is 2.99. However, if the size of the problem is increased to medium size (e.g., 10 seat rows), the solver can no longer find the solution in a reasonable time (more than 10 hours~\citep{Michael_Majid_TRC_2021}). 

In the exemplary use case (Fig.~\ref{fig:small4rows}) with 4 rows, we can see three different parts as follows. 
\begin{itemize}
    \item[] \textbf{Part 1} - Passenger who sits in seats row~1 and column~F (1F) receive the normalized storing shedding rate from passenger 2F when storing two bags.
    \item[] \textbf{Part 2} - Passengers who sit in seats 1A, 4A, and 4F receive the normalized storing shedding rate from passengers 1B, 3A, and 4E, respectively.
    \item[] \textbf{Part 3} - Passengers who sit in seats 1F and 2F receive the normalized storing shedding rate from passenger 1D at the same time.
\end{itemize}

After these three parts, we can calculate the optimal values of received store shedding rates of these passengers; Passenger 1A and 4F (two bags): 0.0598; passenger 1F (one bag): 0.0537; passenger 2F (two bags): 0.0414; passenger 3A and 4A (one bag): 0.009. Also, in the last part of Fig.~\ref{fig:small4rows}, after the boarding process, we can calculate the flight shedding rates based on optimal seat allocation. For example, the sum of normalized flight shedding rates for passenger 1A is 0.207 and for passenger 1D it is 0.141.

As the seat layout problem is a kind of NP-hard problem, we design a novel genetic algorithm (GA) to solve the model for the real size problem, such as an Airbus A320 seat layout with 29 seat rows ($i=29$) and 6 seats per row ($j=6$). The seat layout problem was solved by meta-heuristic algorithms such as GA \citep{van_den_briel_america_2005, majid_soolaki_2012}. Regarding the ability of GA to solve the integer linear programming we use it to find optimal/near-optimal solutions for real size problems. A computer with the specifications of the Core i7- 10610U CPU, 1.80 GHz, 32 GB Ram, and Matlab R2013 are used to run the GA code. 

\subsection{Chromosome structure}
To create the initial generation, first, we define the solutions of chromosome structure as follows.

\vbox{
    \begin{center}
    $
      C=
      \left[ {\begin{array}{cccccc}
       y_{1,1} & y_{1,2} & y_{1,3} & y_{1,4} & y_{1,5} & y_{1,6}\\
       y_{2,1} & y_{2,2} & y_{2,3} & y_{2,4} & y_{2,5} & y_{2,6}\\
       ... & ... & ...& ... & ... & ... \\
       y_{29,1} & y_{29,2} & y_{29,3} & y_{29,4} & y_{29,5} & y_{29,6}\\  
      \end{array} } \right]
    $ \\
    \ \\
    $
      \text{with } y_{i,j}=b \text{ if } x_{i,j,b}=1, \text{otherwise zero}
    $
    \end{center}
}

The arrays of the matrix take values of 0, 1, 2, or 3. If seat ($i$, $j$) is not assigned to any passengers, we consider $y_{i,j}=0$. If there is a passenger with no bag, we consider $y_{i,j}=1$. If there is a passenger with a bag or 2 bags, we consider $y_{i,j}=2$ and $y_{i,j}=3$, respectively. The fitness function is defined based on the original objective function. When developing a first generation and new solutions for the next generations, generated solutions are always feasible. GA operators such as selection, crossover, mutation, and migration generate the next generation from the current generation at each stage.

\subsection{Mutation operators}
The roulette wheel as our selection operator guarantees that the solution with a better (i.e., lower) fitness function has a better chance to select for the following GA operators such as mutation and crossover. Five different operators are designed to implement mutation, as follows. To create the offspring from the parents, we use the cross\-over operator. Firstly, we select a solution and then create a new solution from it. In the first type, we select a non-zeros array and then substitute it with a zero array. We randomly select two non-zero arrays and then change their values in the second mutation operator. In the third and fourth types, we change the arrays of two random rows and columns. Also, we change the location of a random array with a value of 2 or 3 to a zero-window seat location (i.e., column 1 or 6).\\

\vbox{
    \centering
    \hspace{0.5cm} Selected solution \hspace{1.2cm} First new solution\\%
    \quad\\%
    \noindent \resizebox{.48\textwidth}{!}{%
    $%
      \left[ {\begin{array}{cccccc}%
       \tcw{0} & \tcw{3} & \tcw{2} & \tcw{1} & \tcw{0} & \tcw{1}\\%
       \tcw{1} & 0 & 3 & 1 & 0 & 0\\%
       \tcw{...} & ... & ...& ... & ... & ... \\%
       \tcw{2} & 0 & 0 & 3 & 0 & 2\\  %
      \end{array} } \right]%
      \left[ {\begin{array}{cccccc}%
       \textcircled{3} & \textcircled{0} & \tcw{2} & \tcw{1} & \tcw{0} & \tcw{1}\\%
       \tcw{1} & 0 & 3 & 1 & 0 & 0\\%
       \tcw{...} & ... & ...& ... & ... & ... \\%
       \tcw{2} & 0 & 0 & 3 & 0 & 2\\%
      \end{array} }\right]%
    $}%
}


\vbox{
    \begin{center}
    \hspace{0.55cm} Second solution \hspace{1.3cm} Third new solution\\
    \quad\\
    \noindent \resizebox{.48\textwidth}{!}{%
    $%
      \left[ {\begin{array}{cccccc}
       \tcw{0} & \tcw{3} & \textcircled{1} & \textcircled{2} & \tcw{0} & \tcw{1}\\
       \tcw{1} & 0 & 3 & 1 & 0 & 0\\
       \tcw{...} & ... & ...& ... & ... & ... \\
       \tcw{2} & 0 & 0 & 3 & 0 & 2\\  
      \end{array} }\right]
       \left[ {\begin{array}{cccccc}
       \textcircled{1} & \textcircled{0} & \textcircled{3} & \textcircled{1} & \textcircled{0} & \textcircled{0}\\
       \textcircled{0}  & \textcircled{3} & \textcircled{2} & \textcircled{1} & \textcircled{0} & \textcircled{1}\\
       \tcw{...} & ... & ...& ... & ... & ... \\
       \tcw{2} & 0 & 0 & 3 & 0 & 2\\  
      \end{array} } \right]
    $}%
    \end{center}
}


\vbox{
    \begin{center}
        \hspace{0.6cm} Fourth solution \hspace{1.4cm} Fifth new solution\\
        \quad\\
        \resizebox{.48\textwidth}{!}{%
        $%
          \left[ {\begin{array}{cccccc}
           \textcircled{3} & \textcircled{0} & \tcw{2} & \tcw{1} & \tcw{0} & \tcw{1}\\
           \textcircled{0} & \textcircled{1} & 3 & 1 & 0 & 0\\
           \tcw{...} & ... & ...& ... & ... & ... \\
           \textcircled{0} & \textcircled{2} & 0 & 3 & 0 & 2\\  
          \end{array} }\right]
          \left[ {\begin{array}{cccccc}
           \tcw{0} & \tcw{3} & \tcw{2} & \tcw{1} & \tcw{0} & \tcw{1}\\
           \tcw{1} & 0 & 3 & 1 & 0 & 0\\
           \tcw{...} & ... & ...& ... & ... & ... \\
           \tcw{2} & 0 & 0 & \textcircled{2} & 0 & \textcircled{3}\\  
          \end{array} }\right]%
        $}%
    \end{center}
}
\normalsize
\vspace{0.2cm}

\subsection{Crossover operator}
We divide the rows of the aircraft into four different parts: from row 1 to 7, from row 8 to 14, from row 15 to 21, and from row 22 to 29. The first offspring receives their genes related to parts 1 and 3 from the first parent and parts 2 and 4 from the second parent. Similarly, the second offspring gets the 1 and 3 from the second parent and so on.

\vspace{0.5cm}
\vbox{
    \begin{center}
        \noindent 
    Row \quad\quad\quad Parent 1 \quad\quad\quad\quad\quad\quad\quad Parent 2\quad\quad\quad\quad\\
        \quad\quad\\
        \resizebox{.48\textwidth}{!}{%
        $%
          \left[ {\begin{array}{l}
           \tcw{1} \\
           \tcw{...}\\
           \tcw{7}\\
           \tcw{8}\\
           \tcw{...}\\
           \tcw{14}\\
           \tcw{15}\\
           \tcw{...}\\
           \tcw{21}\\
           \tcw{22}\\
           \tcw{...}\\
           \tcw{29}\\
          \end{array} }\right]
          \left[ {\begin{array}{cccccc}
           \tcw{3} & \tcw{0} & \tcw{2} & \tcw{1} & \tcw{0} & \tcw{1}\\
           \tcw{...} & ... & ...& ... & ... & ... \\
           \tcw{2} & 2 & 1 & 1 & 3 & 0\\
           \tcw{0} & 1 & 1 & 1 & 0 & 3\\
           \tcw{...} & ... & ...& ... & ... & ... \\
           \tcw{1} & 3 & 0 & 1 & 0 & 0\\ 
           \tcw{2} & 0 & 0 & 3 & 2 & 2\\
           \tcw{...} & ... & ...& ... & ... & ... \\
           \tcw{2} & 3 & 0 & 3 & 0 & 1\\
           \tcw{0} & 0 & 3 & 1 & 1 & 1\\
           \tcw{...} & ... & ...& ... & ... & ... \\
           \tcw{3} & 1 & 0 & 3 & 1 & 2\\
          \end{array} }\right]
          \left[ {\begin{array}{cccccc}
           \textcircled{1} & \textcircled{3} & \textcircled{0} & \textcircled{1} & \textcircled{2} & \textcircled{1}\\
           \textcircled{...} & \textcircled{...} & \textcircled{...}& \textcircled{...} & \textcircled{...} & \textcircled{...} \\
           \textcircled{1} & \textcircled{1} & \textcircled{3} & \textcircled{0} & \textcircled{1} & \textcircled{3}\\
           \textcircled{2} & \textcircled{0} & \textcircled{2} & \textcircled{0} & \textcircled{1} & \textcircled{2}\\
           \textcircled{...} & \textcircled{...} & \textcircled{...}& \textcircled{...} & \textcircled{...} & \textcircled{...} \\
           \textcircled{0} & \textcircled{2} & \textcircled{0} & \textcircled{1} & \textcircled{0} & \textcircled{3}\\ 
           \textcircled{1} & \textcircled{3} & \textcircled{1} & \textcircled{2} & \textcircled{3} & \textcircled{1}\\
           \textcircled{...} & \textcircled{...} & \textcircled{...}& \textcircled{...} & \textcircled{...} & \textcircled{...} \\
           \textcircled{1} & \textcircled{0} & \textcircled{3} & \textcircled{2} & \textcircled{0} & \textcircled{0}\\
           \textcircled{0} & \textcircled{2} & \textcircled{1} & \textcircled{2} & \textcircled{3} & \textcircled{0}\\
           \textcircled{...} & \textcircled{...} & \textcircled{...}& \textcircled{...} & \textcircled{...} & \textcircled{...} \\
           \textcircled{1} & \textcircled{1} & \textcircled{3} & \textcircled{1} & \textcircled{3} & \textcircled{1}\\
          \end{array} }\right]%
        $}%
    \end{center}
}

\vspace{0.1cm}

\vbox{%
    \noindent %
    \begin{center}%
            \noindent 
    Row \quad\quad Offspring 1 \quad\quad\quad\quad\quad Offspring 2\quad\quad\quad\quad\\
        \quad\quad\\
    \resizebox{.48\textwidth}{!}{%
      $%
      \left[ {\begin{array}{l}
       \tcw{1} \\
       \tcw{...}\\
       \tcw{7}\\
       \tcw{8}\\
       \tcw{...}\\
       \tcw{14}\\
       \tcw{15}\\
       \tcw{...}\\
       \tcw{21}\\
       \tcw{22}\\
       \tcw{...}\\
       \tcw{29}\\
      \end{array} }\right]
      \left[ {\begin{array}{cccccc}
       \tcw{3} & 0 & 2 & 1 & 0 & 1\\
       \tcw{...} & ... & ...& ... & ... & ... \\
       \tcw{2} & 2 & 1 & 1 & 3 & 0\\
       \textcircled{2} & \textcircled{0} & \textcircled{2} & \textcircled{0} & \textcircled{1} & \textcircled{2}\\
       \textcircled{...} & \textcircled{...} & \textcircled{...}& \textcircled{...} & \textcircled{...} & \textcircled{...} \\
       \textcircled{0} & \textcircled{2} & \textcircled{0} & \textcircled{1} & \textcircled{0} & \textcircled{3}\\ 
       \tcw{2} & 0 & 0 & 3 & 2 & 2\\
       \tcw{...} & ... & ...& ... & ... & ... \\
       \tcw{2} & 3 & 0 & 3 & 0 & 1\\
       \textcircled{0} & \textcircled{2} & \textcircled{1} & \textcircled{2} & \textcircled{3} & \textcircled{0}\\
       \textcircled{...} & \textcircled{...} & \textcircled{...}& \textcircled{...} & \textcircled{...} & \textcircled{...} \\
       \textcircled{1} & \textcircled{1} & \textcircled{3} & \textcircled{1} & \textcircled{3} & \textcircled{1}\\
      \end{array} }\right]
      \left[ \begin{array}{cccccc}
       \textcircled{1} & \textcircled{3} & \textcircled{0} & \textcircled{1} & \textcircled{2} & \textcircled{1}\\
       \textcircled{...} & \textcircled{...} & \textcircled{...}& \textcircled{...} & \textcircled{...} & \textcircled{...} \\
       \textcircled{1} & \textcircled{1} & \textcircled{3} & \textcircled{0} & \textcircled{1} & \textcircled{3}\\
       \tcw{0} & 1 & 1 & 1 & 0 & 3\\
       \tcw{...} & ... & ...& ... & ... & ... \\
       \tcw{1} & 3 & 0 & 1 & 0 & 0\\ 
       \textcircled{1} & \textcircled{3} & \textcircled{1} & \textcircled{2} & \textcircled{3} & \textcircled{1}\\
       \textcircled{...} & \textcircled{...} & \textcircled{...}& \textcircled{...} & \textcircled{...} & \textcircled{...} \\
       \textcircled{1} & \textcircled{0} & \textcircled{3} & \textcircled{2} & \textcircled{0} & \textcircled{0}\\
       \tcw{0} & 0 & 3 & 1 & 1 & 1\\
       \tcw{...} & ... & ...& ... & ... & ... \\
       \tcw{3} & 1 & 0 & 3 & 1 & 2\\
      \end{array} \right]%
    $}%
    \end{center}
}

\vspace{0.5cm}


When the first generation is created, all solutions are feasible because the number of passengers with a specific value of bags (e.g., 29 passengers with two bags) is determined. Each solution is a matrix of size 29 by 6 (number of rows is 29 and number of columns is 6). The arrays of the matrix could take the values of 0 (empty seat), 1 (passenger without bag), 2 (a passenger with a bag), or 3 (a passenger with two bags). For the first generation, we start with a zero matrix, and then we select 29 zero arrays and substitute them with the value of 3, which means they have 2 bags and so on. Each solution in the first generation of GA has 29 arrays of one, 58 arrays of two, 29 arrays of three, and 58 arrays of zero when the GA is run for the second scenario. The only important constraint in the GA is the number of these values to have feasible solutions in the first generation. The new solutions for the next generations are always feasible. Two important operators are implemented here, mutation and crossover. In mutation, we always create a feasible solution because we change the location of two non-zero arrays or select a non-zero array to put it in a zero array and then the first array takes zero. In the crossover operator, solutions take their arrays from two parents. Based on this operator, infeasible solutions might be generated. We use a strategy to modify these solutions to have always feasible solutions. For example, instead of having a solution with 29 arrays of one, 58 arrays of two, 29 arrays of three, and 58 arrays of zero, after implementing the crossover, we have a solution with 28 arrays of one, 59 arrays of two, 27 arrays of three and 60 arrays of zero. We need to have another one in the solution. Thus, we randomly select one of the zeros and assign it to one. As a result, we have 29 arrays of one, 59 arrays of two, 27 arrays of three, and 59 arrays of zero. The final step is selecting two arrays which are randomly taken 2 and 0 and assigning them the value of three. We reach the main assumption, which is 29 arrays of one, 58 arrays of two, 29 arrays of three, and 58 arrays of zero. The feasible solutions will be generated when creating the next generation from the current one under different operators.

\subsection{Additional operators}
To enable to transfer of a given (low) percentage of the current generation to the next, the operators elitism and migration are implemented. The best solutions based on fitness function are transferred to the next generation under the elitism operator, which guarantees the best solution in each generation is equal to or better than the best solution in the previous generation. The migration operator enhances the population's genes by adding randomly generated sequences, which expands diversity and accelerates the algorithm's convergence. Both operators are defined by a share of the population size. We consider the following parameters of GA: population size~=~200, generations~=~1,000, mutation rate~=~30\%, crossover rate~=~ 50\%, and elitism and migration rates~=~10\%.

\subsection{Application scenarios}
Here we define three strategies in terms of the number of seats assigned to passengers and the number of bags that passengers have as follows.

\begin{figure}[htb!]
    \centering
    \includegraphics[width=.6\textwidth]{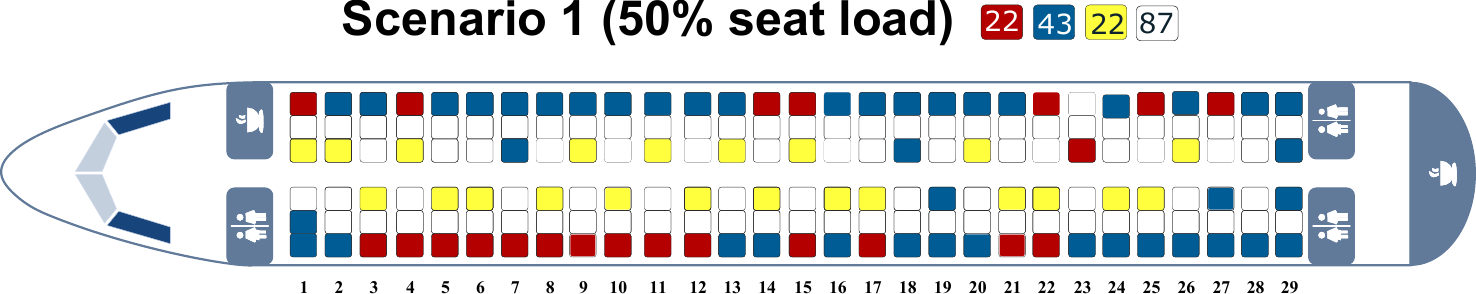}
    \caption{Scenario 1.}
    \label{fig:GA050}
\end{figure}

\paragraph{\textbf{Scenario 1}} Seat load of 50\% which means 87 passengers including 22 passengers without bags, 43 passengers with a bag, 22 passengers with two bags, and 87 empty seats  (Fig.~\ref{fig:GA050}).

\begin{figure}[htb!]
    \centering
    \includegraphics[width=.6\textwidth]{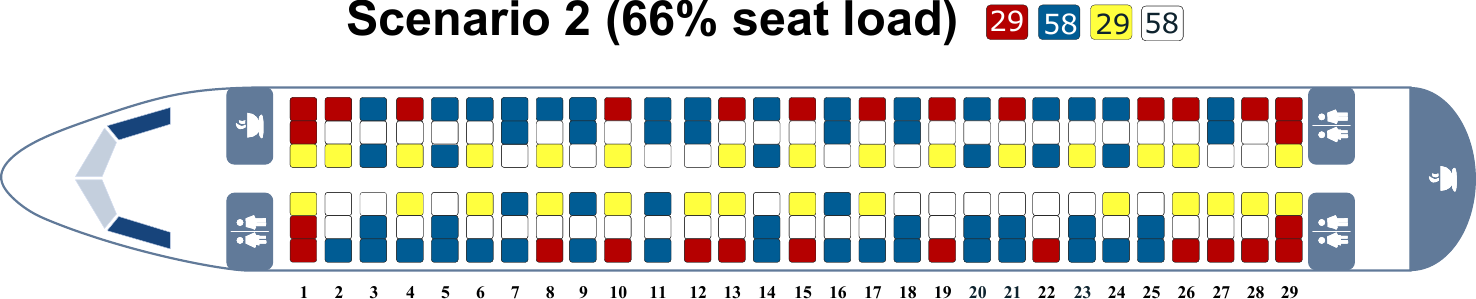}
    \caption{Scenario 2.}
    \label{fig:GA066}
\end{figure}

\paragraph{\textbf{Scenario 2}} Seat load of 66\% which means 116 passengers including 29 passengers without bags, 58 passengers with a bag, 29 passengers with two bags, and 58 empty seats (Fig.~\ref{fig:GA066}).

\begin{figure}[htb!]
    \centering
    \includegraphics[width=.6\textwidth]{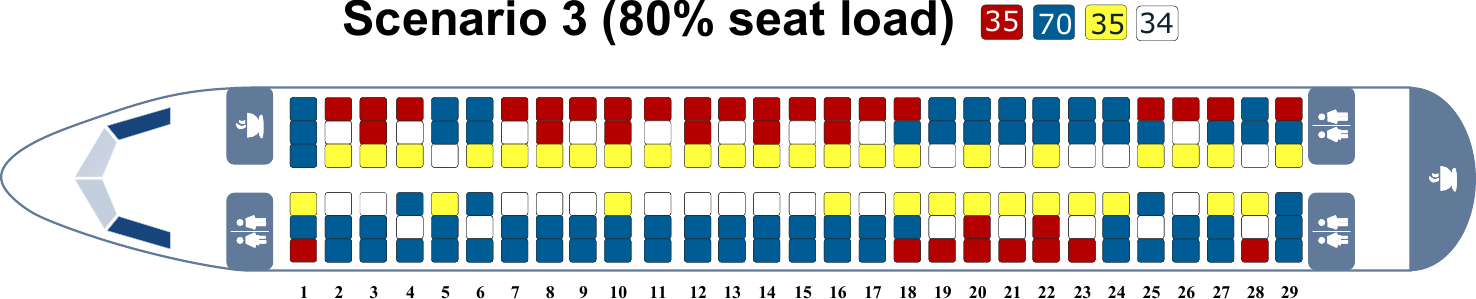}
    \caption{Scenario 3.}
    \label{fig:GA080}
\end{figure}

\paragraph{\textbf{Scenario 3}} Seat load of 80\% which means 140 passengers including 35 passengers without bags, 70 passengers with a bag, 35 passengers with two bags, and 34 empty seats (Fig.~\ref{fig:GA080}).


The GA results in Fig.~\ref{fig:GA080} best highlights how our approach works. Passengers without a bag are seated in the aisle seats (i.e., columns C and D), so they will not stay long in the aisle for baggage storing. Consequently, passengers with one bag and two bags sit in middle/aisle seats and while they stow their luggage, only a few seats in front of them are occupied by other passengers. Both configurations reduce the transmission risk for all already seated passengers. Further, the algorithm results in a balanced distribution of compartment utilization per side (left/ right of the aisle and per row (front/ middle/ rear part of the aircraft). Considering the corresponding scenario assumptions, a maximum of 4 pieces of hand luggage are stored in the respective overhead compartment.

\begin{figure}[htb!]
    \centering
    \includegraphics[width=.75\textwidth]{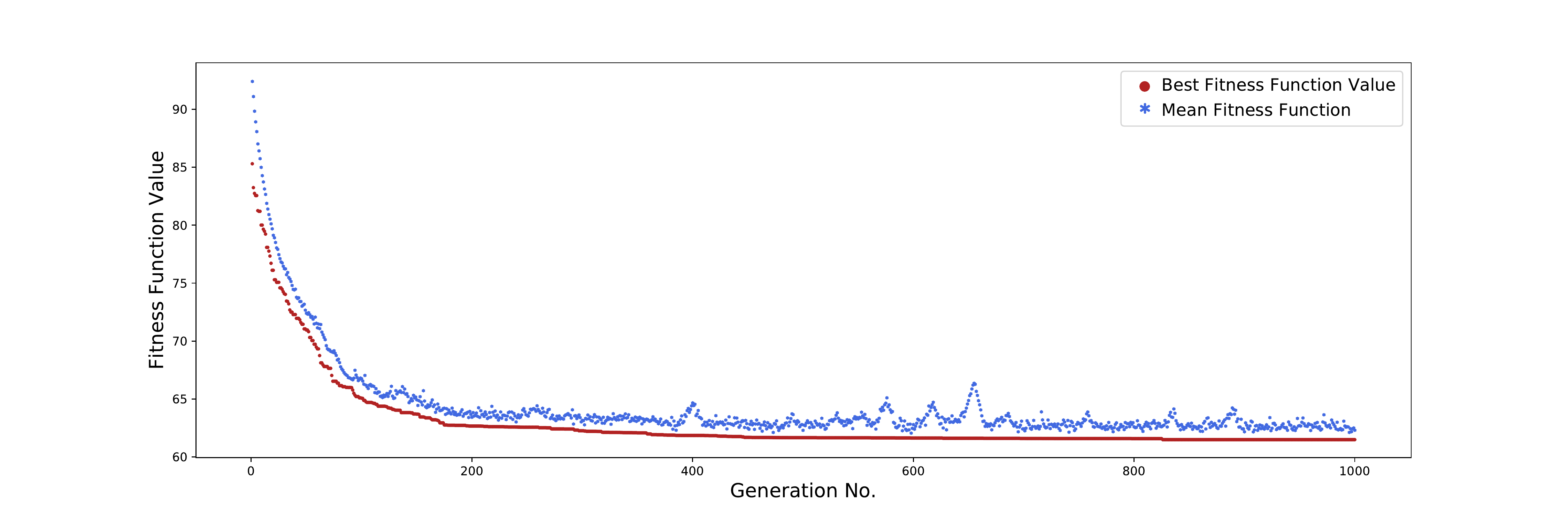}
    \caption{The evolutionary diagram for the Second Scenario}
    \label{fig:GAEvol.}
\end{figure}

The run time of each scenario is lower than 5~minutes. The best fitness values for the three scenarios are 26.3963, 61.4913, and 104.2342, respectively. For example, the run time of the second scenario is 74.4~seconds, and fitness values for both best and average fitness function for each generation are shown in Fig.~\ref{fig:GAEvol.}. 
The proposed GA converges to the near/optimal solution in the 826$^{th}$ generation.

In the next step, we will use the derived seat allocations as input for a boarding simulation. Depending on different boarding sequences, the impact of increasingly utilized overhead compartments on passenger boarding time and virus transmission risk will be evaluated.

\section{Agent-based simulation model}
\label{sec:agentSimulation}
The individual movement behavior of passengers in the aircraft cabin is modeled by a validated, stochastic cellular automaton approach, which covers short (e.g., avoid collisions, group behavior) and long-range interactions (e.g., tactical way\-finding), as well as the stochastic nature of passenger movements~\citep{_Schultz2010,schultz_implementation_2018}. This cellular automaton model is based on an individual transition matrix, which contains transition probabilities for a passenger to move to adjacent positions around the current location~\citep{_Schultz2013d}.

\subsection{Grid-based transition approach for passenger movements}
\begin{figure}[htb!]
    \centering
    \includegraphics[width=.75\textwidth]{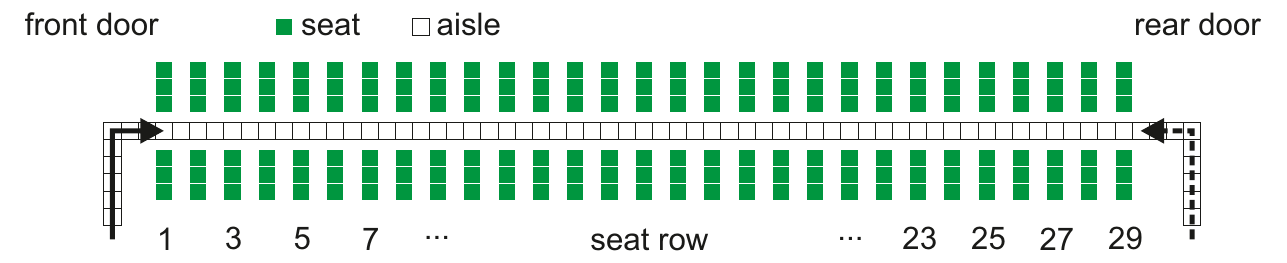}
    \caption{Grid-based aircraft model with 29 seat rows and 6 seats per row (reference layout for single-aisle, narrow-body configurations). Layout shows one door in use for disembarkation.}
    \label{fig:a320_grid}
\end{figure}

The implemented cellular automaton model considers operational conditions of aircraft and airlines (e.g., seat load factor, conformance to the boarding procedure) as well as the non-deterministic nature of the underlying passenger processes (e.g., hand luggage storage) and was calibrated with data from the field \citep{schultz_fieldTrial_2018}. The cellular automaton for aircraft boarding and disembarkation is based on a regular grid (Fig.~\ref{fig:a320_grid}), which consists of equal cells with a size of 0.4 x 0.4 m, where a cell can either be empty or contain exactly one passenger. Passengers can only move one cell per timestep or must stop if the cell in the direction of movement is occupied. 

The boarding progress consists of a simple set of rules for the passenger movement: (a) enter the aircraft at the assigned door (based on the current boarding scenario), (b) move forward from cell to cell along the aisle until reaching the assigned seat row, and (c) store the luggage (aisle is blocked for other passengers) and take the seat. The storage time for the hand luggage depends on the individual number of hand luggage items. The seating process depends on the constellation of already used seats in the corresponding row. The agents are sequenced concerning the active boarding strategy. From this sequence, a given percentage of agents are taken out of the sequence (non-conforming behavior) and inserted into a position, which contradicts the strategy (e.g., inserted into a different boarding block). 

The maximum, free walking speed in the aisle is 0.8~m/s \citep{_Schultz2018b}, so a simulation timestep is 0.5~s. In each simulation step, the list of passengers to be updated is randomly shuffled to emulate a parallel update behavior for the discrete time-dynamics (random-sequential update) \citep{_Schultz2010,_Schultz2013d}. The boarding time is defined as the time between the first passenger entering the aircraft and the time the last passenger is seated. Each boarding scenario is simulated 125,000 times, to achieve statistically relevant results defined by the average boarding time. Further details regarding the general stochastic model, parameter setups, and the simulation environment are provided in \citep{schultz_implementation_2018}. 

To minimize the risk of virus transmission, a physical distance between passengers is required. The International Aviation Transport Association demands a distance of at least 1~meter~\citep{IATA_CabinOps_2020} and the Federal Aviation Administration a minimum of 6~feet (2~meters)~\citep{FAA_COVID19_2020}. Considering the cellular automaton model with a grid of 0.4~x~0.4~m cells, and to maintain comparability of our results with preliminary studies \citep{schultz_evaluation_2020,schultz_future_2020,Michael_Majid_TRC_2021}, the minimum physical distance was set to 1.6~m (4~cells). At this point, we assume that passengers are informed that 1.6~m corresponds to the distance of 2 seat rows, which offers proper visual guidance. To assess the overall transmission risk of a given scenario, in every simulation run one passenger is randomly marked as infected. During each timestamp, the individual transmission risk is calculated and updated for each passenger based on the continuous transmission risk approach defined in Section~\ref{sec:transmissionrisk}. After completion of a simulation run, all individual risks are summed up. The average values from the 125,000 simulation runs are used as the representative values for the transmission risk of the corresponding scenario. 

The application of different seat assignments and boarding sequences result in different passenger interactions and utilization of overhead compartments within the simulated aircraft cabin environment, thus increasing/decreasing the observed boarding times and transmission risks. Seat assignment and boarding sequence are input parameters for the agent-based simulation and must be defined before a simulation run starts. Even if optimization algorithms are implemented in the simulation environment itself (cf.~\cite{schultz2019machine,Schultz_2017_dynamic_change}, they are not used in this study. While the seat assignment comes from the optimization performed above (see Section~\ref{sec:optresults}), the improved outside-in boarding sequences were taken from a previous research study~\citep{Michael_Majid_TRC_2021}.

\subsection{Hand luggage and compartment utilization}
\label{sec:compartment_util}

A common modelling approach for the time needed to store hand luggage items in the overhead compartment ($t_{\text{store items}}$) is given with (\ref{eq:bag_nagel}) and was used in several research studies~\citep{2005_Ferrari_Robustness,Audenaert_2009,MILNE2016_104}. The individual number of items $n_\text{pax}$ increases the number of already stored items $n_\text{bin}$, and $t_{\text{store items}}$ is calibrated with the parameter $a$ and scaled with a time dimension $\Delta t$ (e.g., simulation step size). This approach leads to increasingly longer times, both with a higher number of individual items and with a higher number of already stowed luggage. However, the capacity limit of the overhead compartment is not considered.

\begin{equation}
    t_{\text{store items}} = \left( a + \frac{n_{\text{bin}} + n_{\text{pax}}}{2} \ n_{\text{pax}} \right)  \Delta t
    \label{eq:bag_nagel}
\end{equation}

An idea motivated by the Pollaczek-Khintchine equation (\ref{eq:Pollaczek}) allows to overcome this limitation by considering the utilization $\rho = \lambda/\mu$. Eq. (\ref{eq:Pollaczek}) generally describes the average waiting time $t_w$ in a queue with arrival rate $\lambda$, capacity $\mu$, and variance for arrivals $\sigma$~\citep{pollaczek_uber_1930,khintchine_mathematical_1932}. If we are assuming that $\sigma^2$ and $\mu$ are constant, we could simplify the equation to a term which considers the arrival rate $\lambda$ (in our case, the number of individual items $n_\text{pax}$), utilization of overhead compartment $\rho$ in percent, and the calibration parameter $a$.

\begin{equation}
    t_w = \frac{\lambda \left( \sigma^2 + \frac{1}{\mu^2} \right)}{ 2 \left( 1 - \rho \right)} 
    = \frac{a \ \lambda}{ 1 - \rho}
    \label{eq:Pollaczek}
\end{equation}
 
Like (\ref{eq:bag_nagel}), $t_{\text{store items}}$ increase in (\ref{eq:Pollaczek}) with a higher number of individual items but also with a higher utilization of the compartment. When the capacity is almost reached, the corresponding strong increase in storage times is now effectively considered, but $\rho=1$ results in $t_\text{store items}=\infty$. We set $\rho_\text{max} = 0.9$ for the calculation and further assume 6  as maximum capacity $n_\text{max}$ for the overhead compartment, with one bigger and one smaller item for each passenger. Considering the previously described assumptions, (\ref{eq:bagTimeCapacity}) is derived. It may be noted that $\sum^i n_\text{pax, i}$ considers the utilization of the overhead compartment \emph{including} the items the current passenger $i$ stores.

\begin{equation}
    t_{\text{store items, i}} = \frac{a \ n_\text{pax, i}}{1 - min \left( \rho_{max} ,\frac{\sum^i n_\text{pax, i}}{{n_\text{max}}} \right) }
    \label{eq:bagTimeCapacity}
\end{equation}

\begin{figure}[htb!]
    \centering
    \includegraphics[width=.6\textwidth]{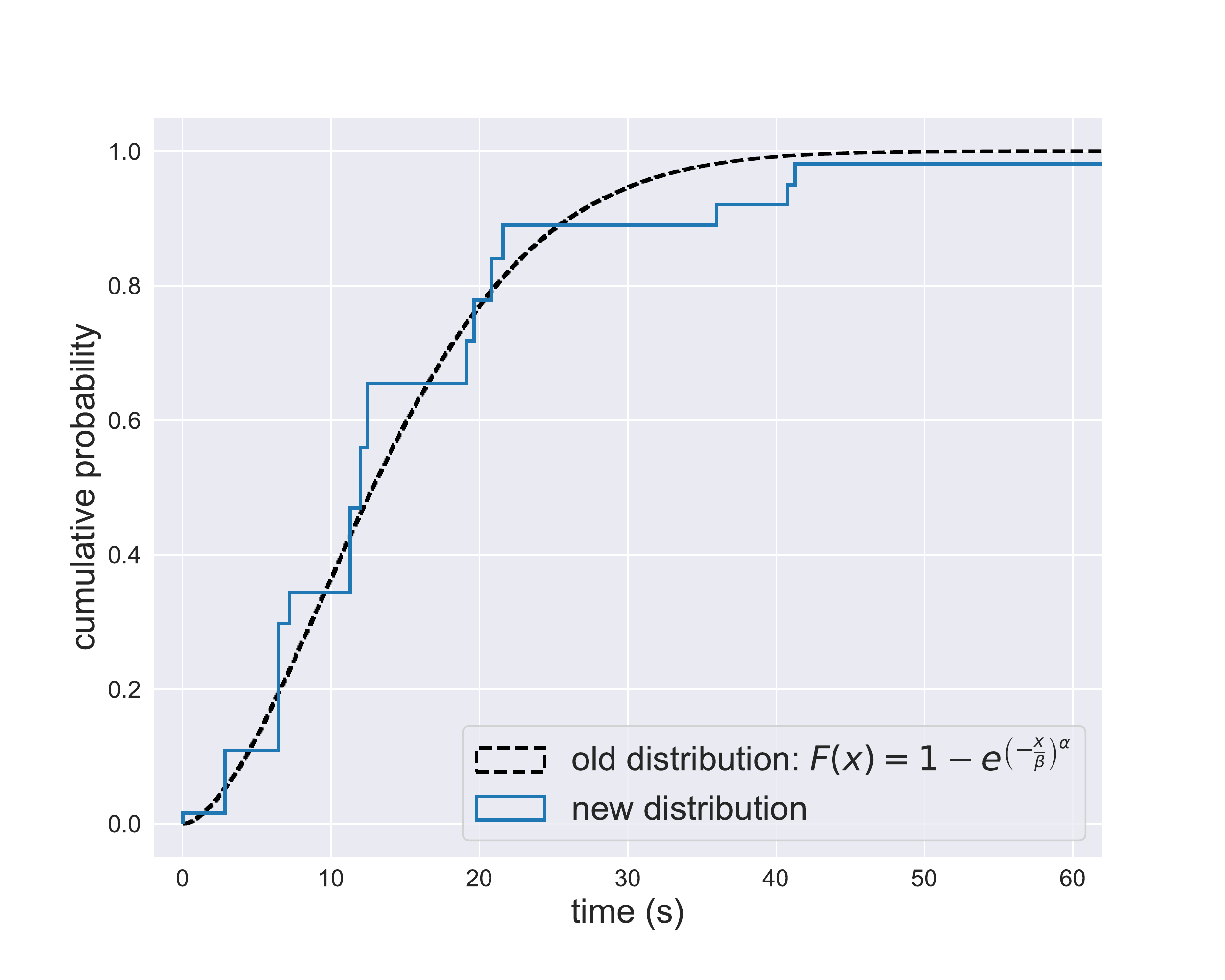}
    \caption{Times needed to store hand luggage items with increasing utilization of the overhead compartment: fitting of new time distribution regarding to a field trials reference.}
    \label{fig:bagTimeStorage01}
\end{figure}

The passengers in the scenario populations possess 0, 1, or 2 hand luggage items with a probability of 25\%, 50\%, and 25\% respectively. A general distribution of times for storing the hand luggage is derived from field trials and fitted to a Weibull distribution with the parameters $\alpha = 1.7$ and $\beta = 16~s$ (Fig.~\ref{fig:bagTimeStorage01}, black dashed line \citep{schultz_fieldTrial_2018}).

\begin{figure}[htb!]
    \centering
    \includegraphics[width=.6\textwidth]{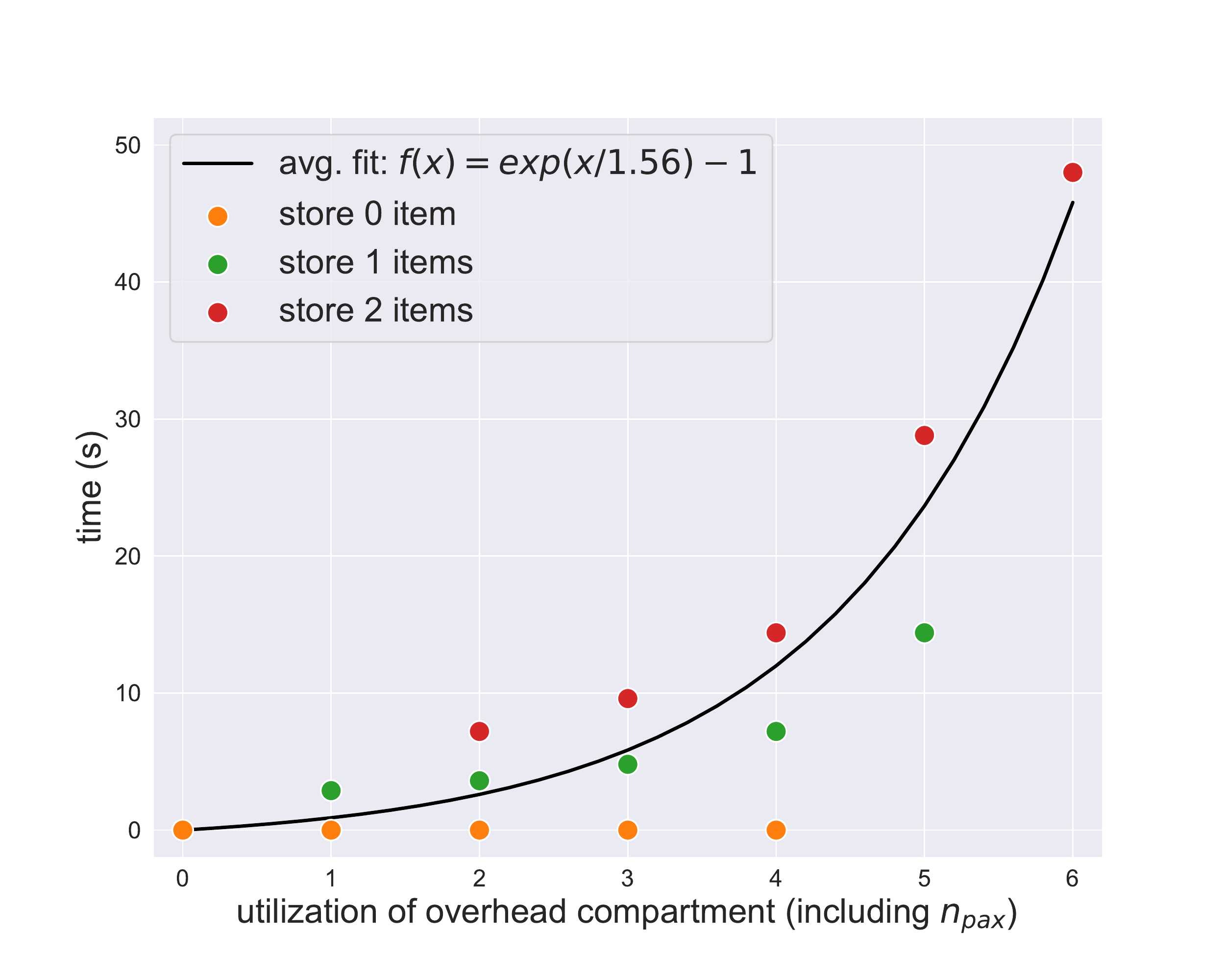}
    \caption{Times needed to store hand luggage items with increasing utilization of the overhead compartment: individual time to store hand luggage items depending on the number of individual items and final utilization of compartment. The black line illustrates the average progressive time growth.}
    \label{fig:bagTimeStorage02}
\end{figure}

This distribution is used as reference to calculate the parameter $a = 2.4$~\emph{s}/\emph{bag} from (\ref{eq:bagTimeCapacity}) using the baggage item distribution in the assumed population. For example, the time for storing two items in the overhead compartment increases from 7.2~s, 9.6~s, 14.4~s, 28.8~s, to 48~s when the overhead compartment already contains 0, 1, 2, 3, and 4 items, respectively (Fig.~\ref{fig:bagTimeStorage02}). 

Consequently, the maximum time (storing six items in the compartment) is 7.2~s + 14.4~s + 48~s = 69.6~s, which happens if all passengers from one side of a seat row (three seats) carrying two hand luggage items (probability of 1.56\%).

\subsection{Implementation and evaluation of scenarios}
\begin{figure}[htb!]
    \centering
    \includegraphics[width=.7\textwidth]{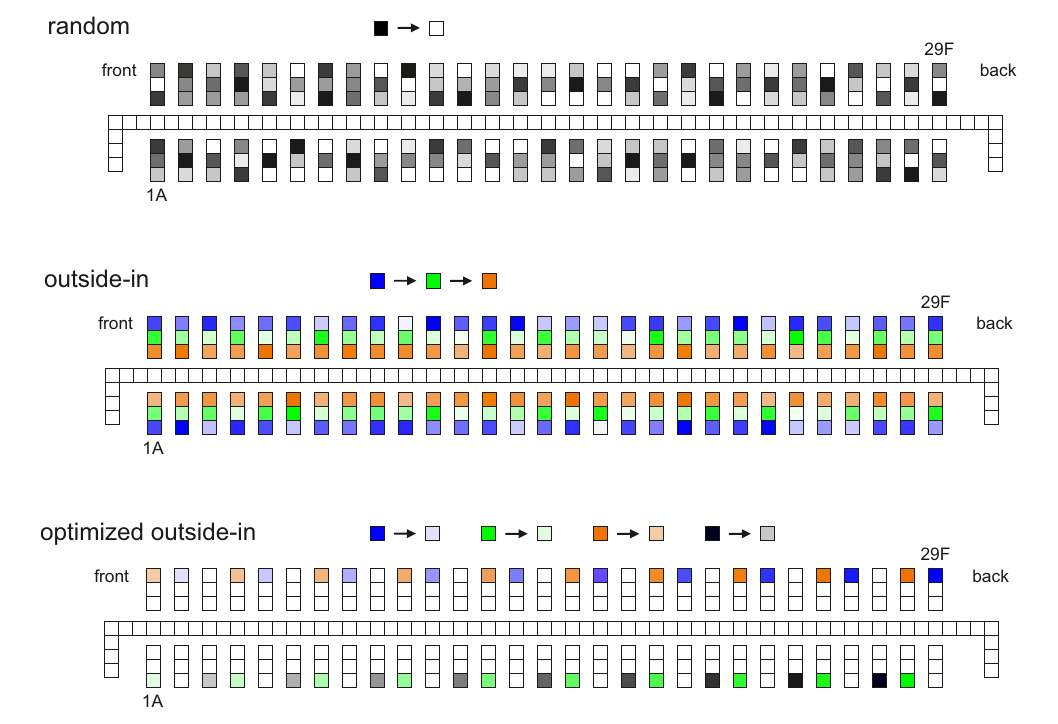}
    \caption{Three implemented boarding strategies, with a random, outside-in, and optimized outside-in sequence. Boarding is based on the designated color sequence, with dark seats being used first and light seats last.}
    \label{fig:boardingProcedures}
\end{figure}

For the assessment of the scenarios, a random boarding procedure (randomly distributed seats and no specific boarding order) was implemented as a reference case. This allows evaluating both the effect of the optimized seat allocation and the additional improvement due to an optimized boarding sequence. In Fig.~\ref{fig:boardingProcedures} (top), a random boarding sequence is shown, where darkly marked seats are filled first, and lightly marked seats are filled last.

In the (common) \emph{outside-in} strategy, passengers are divided into three groups according to their seating position. Passengers with window seats board first and those with aisle seats board last (Fig.~\ref{fig:boardingProcedures}, center). This simple three-group approach does not consider any particular order of passengers within a group. This approach reduces the negative impact of interactions in the seat row when passengers must leave the seat row to allow following passengers to take their seats located closer to the window.

In the proposed \emph{optimized outside-in} strategy, passengers are additionally sorted in each group, corresponding to a staggered approach from the back to the front of the aircraft~\citep{Michael_Majid_TRC_2021}. Taken the first boarding group (window seats) as an example, the list of seats from the back to the front (29F, 28F, .., 1F) is separated into sublists (29F, 26F, 23F, 20F, 17F, 14F, 11F, 8F, 5F, 2F), (28F, 25F, 22F, 19F, 16F, 13F, 10F, 7F, 4F, 1F), (..). These partial lists must account for the requirement of a physical distance between passengers, which we have realized in the implementation by spacing them 3 rows apart. This additional sorting of passengers into groups reduces the likelihood of passengers passing each other, e.g., a passenger from row~20 must wait for the passenger in front of him/her who has a seat in row~10. Also, the left and right sides of the cabin are boarded in an alternating and staggered pattern to further reduce negative passenger interactions (Fig.~\ref{fig:boardingProcedures}, below). 

Finally, there are two seat assignments, optimized layout and randomly distributed seats, and three passenger boarding sequences, sorted by ascending complexity: random, outside-in, and optimized outside-in. The results of the passenger boarding simulation are shown in Table~\ref{tab:simulation_results}. As mentioned earlier, each scenario is simulated 125,000 times, with a new random boarding sequence and seat allocation in the corresponding scenarios provided for each simulation run. The average value for the overall transmission risk, the sum over all individual transmission risk, and the average boarding time are reported in Table~\ref{tab:simulation_results}. For each of the three scenarios, a reference implementation is simulated with a random seat layout and random boarding sequence, with the average boarding time set as a benchmark at 100\%. Regarding this benchmark, it can be seen that both the implementation of the outside-in procedure and the optimized seat allocation led to a reduction in boarding time and a reduction in the transmission risk. 

\begin{table*}[htb!]
    \footnotesize
    \centering
    \begin{tabular}{lllrr}
        \hline
        \bfseries Scenario& \bfseries Seat allocation & \bfseries Boarding strategy & \bfseries Average trans- & \bfseries Average boarding\\
            &   &   & \bfseries mission risk & \bfseries time (\%)\\
        \hline
        1 (50\% seat load)    &optimized layout&optimized outside-in                       &0.00    &46\\
                &               &outside-in  &0.00    &66\\
                &               &random sequence                            &0.01    &88\\
        \multicolumn{1}{r}{references:}       &random seats   & outside-in &0.00    &79\\
                                            &   &random sequence                &0.02    &100\\
        \hline
        2 (66\% seat load)&optimized layout&optimized outside-in       &0.00    &45\\
        &&outside-in &0.00   &61\\
        &&random sequence                               &0.01    &91\\
        \multicolumn{1}{r}{references:}       &random seats   & outside-in    &0.00    &76\\
        &&random sequence      &0.02    &100\\
        \hline
        3 (80\% seat load)&optimized layout&optimized outside-in   &0.00    &41\\
         & &outside-in &0.00    &59\\
         & &random sequence                             &0.02    &97\\
         \multicolumn{1}{r}{references:}       &random seats   & outside-in   &0.00    &74\\
         &&random sequence     &0.02     &100\\
         \hline
    \end{tabular}
    \caption{Simulation of appropriate boarding strategies for the defined scenarios. For each scenario, combinations of (non) optimized seat allocation and (non) optimized boarding sequence are analyzed.}
    \label{tab:simulation_results}
\end{table*}

As previous research has shown, the introduction of an outside-in sequence speeds up passenger boarding even without the implementation of appropriate seat layouts. Thus, the boarding time is reduced by 21\%, 24\%, and 26\% for scenarios 1, 2, and 3, respectively (random seats, random sequence to random seats, outside-in). In this case, the transmission risk is reduced from 0.02 to 0. If only the optimized seat layout is applied, considering a random boarding sequence, the boarding time is reduced by 12\%, 9\%, and 3\% for these scenarios. The transmission risk is reduced by 50\% to 0.01. A combined approach of using the common outside-in and the newly developed optimized seat layout results in 34\%, 39\%, and 41\% boarding time reduction, accompanied by a transmission risk of 0. As the seat load factor increases from scenario 1 to 3, the combined implementation leads to even greater improvements at realistic load factors. If in a final step, the order of passengers in each group is also optimized (optimized outside-in), boarding time can be further improved by up to 20\%.

\subsection{Compartment utilization model}
The application of the developed compartment utilization approach (Section~\ref{sec:compartment_util}) shows that, if the increased storage time at higher utilization is not considered, boarding time reduction is overestimated by 30\% on average (relative to a boarding time of the corresponding random boarding case). This shows quite clearly how much the results in this case depend on the modeling approach chosen. The general error of the applied agent-based model was determined to be about 5\%~\citep{schultz_fieldTrial_2018}, so overestimation is significant and cannot be considered a general model uncertainty. If the achieved results are compared by absolute values, the overestimation of boarding time reduction with the old approach (no compartment utilization) is 16\% at average, but still significant. It can therefore be assumed that models that do not consider the degree of utilization of the baggage compartment tend to overestimate the impact of new procedure designs.

\section{Discussion and outlook}
\label{sec:discussion}
To handle the aircraft boarding problem under COVID-19 requirements, we researched the combination of optimization and simulation approaches providing optimized seat allocations and boarding sequences~\citep{Michael_Majid_TRC_2021, Michael_Majid_TransportB_2021, Michael_Majid_ATM_2021}. In these studies, we implemented shedding rates~\citep{schultz_evaluation_2020} to calculate a transmission risk indicator, addressing the potential spread of viruses in the aircraft cabin. We had not considered the utilization of the overhead compartment until now, however, in smaller experiments we were able to determine that this aspect has a significant effect on passenger boarding efficiency, particularly under COVID-19 requirements.

We started our approach with the commonly used outside-in boarding strategy since this strategy already reduces the passenger interactions in the seat rows. We see baggage storage as a time-consuming, physical activity, which leads to a higher risk of virus transmission in the aircraft cabin. Therefore, we first develop a new mathematical model to calculate two types of shedding rates, namely when passengers are seated and when they store their luggage. As the boarding is an NP-hard type, we design a genetic algorithm to solve the problem. Three operational scenarios with a seat load of 50\%, 66\%, and 80\% are defined and the resulting seat allocations are used as input for a stochastic agent-based model to evaluate the average boarding time and transmission risk. The stochastic agent-based simulation demonstrates that the provided seat allocations result in a reduction in boarding times and transmissions risk, even when no specific boarding sequence is provided. However, if the recommended outside-in boarding is implemented and the boarding sequence is additionally optimized, the boarding time can be significantly reduced by more than 50\% compared to the random reference and by more than 30\% compared to the outside-in reference. In these cases, the transmission risk has remained at the lowest level.

Our research results indicate that the process of baggage storage in the aircraft cabin has a significant impact on the process efficiency in terms of boarding time minimization and transmission risk mitigation. At the same time, there is still a high potential for optimization and a need for implementation. The next steps could be to determine the expected number of hand luggage items in real-time as well as the utilization status of the overhead bins during boarding. In addition to pre-operation optimization, this could also lead to dynamic control of the boarding sequence during operations.

The COVID-19 epidemic has drastically reduced the number of passengers transported. The resulting low seat load factors have mitigated the negative impact of legal requirements (e.g., physical distances) on passenger handling processes (e.g., boarding times) to a certain degree. The planned aircraft turnaround times were met due to the reduced number of passengers. With the expected normalization of the air traffic sector, it will no longer be possible to comply with these times. Assuming that pandemic requirements are an obligatory task to be managed by airlines and airport operators, intelligent and operationally relevant solutions must be provided. Passengers should no longer simply board an aircraft, but be guided through an adaptive, optimized layout and boarding process. Digitization already offers an appropriate technological basis for the exchange of information between passengers and operators, e.g., seat location, number of hand luggage items, group size, or dynamic position data. With our concept, new digital approaches could be realized not only to offer (location-based) services for passengers but also to implement sustainable improvements in passenger handling.

\reftitle{References}
\bibliography{bib}

\end{document}